\documentclass{jfm}
\usepackage{amsmath}
\usepackage{hyperref}
\usepackage{graphicx}
\usepackage{amsfonts}
\usepackage{amssymb}
\usepackage{xcolor}
\usepackage{bm}
\usepackage{soul}
\usepackage[version=4]{mhchem}
\usepackage{subcaption}
\usepackage{float}
\usepackage{cancel}
\usepackage{tikz}

\newcommand{\PP}{\mathcal{P}}

\shorttitle{Lattice Boltzmann model for reactive mixtures in porous media}
\shortauthor{N. Sawant and I. V. Karlin}

\title{{Mean field} lattice Boltzmann model for reactive mixtures in porous media}

\author{N. Sawant\aff{1}
 \and I. V. Karlin\aff{1}\corresp{\email{ikarlin@ethz.ch}}}

\affiliation{\aff{1}Department of Mechanical and Process Engineering, ETH Zurich, 8092 Zurich, Switzerland}

\begin{document}

\maketitle

\begin{abstract}
A new lattice Boltzmann model (LBM) is presented to describe chemically reacting multicomponent fluid flow in homogenised porous media. 
In this work, towards further generalizing the multicomponent reactive lattice Boltzmann model, we propose a formulation which is capable of performing reactive multicomponent flow computation in porous media at the representative elementary volume (REV) scale. To that end, the submodel responsible for interspecies diffusion has been upgraded to include Knudsen diffusion, whereas the kinetic equations for the species, the momentum and the energy have been rewritten to accommodate the effects of volume fraction of a porous media though careful choice of the equilibrium distribution functions. Verification of the mesoscale kinetic system of equations by a Chapman--Enskog analysis reveals that at the macroscopic scale, the homogenized Navier--Stokes equations for compressible multicomponent reactive flows are recovered. The Dusty Gas Model (DGM) capability hence formulated is validated over a wide pressure range by comparison of experimental flow rates of component species counter diffusing through capillary tubes. Next, for developing a capability to compute heterogeneous reactions, source terms for maintaining energy and mass balance across the fluid phase species and the surface adsorbed phase species are proposed. The complete model is then used to perform detailed chemistry simulations in porous electrodes of a Solid Oxide Fuel Cell (SOFC), thereby predicting polarization curves which are of practical interest. 
\end{abstract}



\section{Introduction}
\label{sec:intro}

Fluid flow and transport in porous media is ubiquitous, with applications ranging across disciplines such as but not limited to earth science, biology and energy science. Some areas relevant to the times which can benefit from modeling of transport in porous media  include subsurface carbon dioxide or hydrogen storage in geological formations, geothermal power exploration, renal and pulmonary flows, flows in electrochemical energy devices etc. In this work, we develop a lattice Boltzmann model for multicomponent reactive flows with a vision of creating a useful prediction tool for the operation of fuel cells made up of porous electrodes.

Let us take a brief overview of the literature surrounding simulation of fuel cells, especially the solid oxide fuel cells (SOFCs). As with any other area of modeling and simulation, to suit an expected level of accuracy, different models varying in complexity of the science as well as dimensions exist. One of the most utilitarian and well validated model is the one-dimensional button-cell model developed by \cite{decaluwe_importance_2008} for the purpose of exploring the influence of anode microstructure on SOFC performance. The model computes coupled dusty gas model (DGM) and elementary electrochemical kinetics in a porous nickel–yttria stabilized zirconia (YSZ) cermet anode, a dense YSZ electrolyte membrane, and a composite lanthanum strontium manganite (LSM)–YSZ cathode. The effects on overpotential of microstructural parameters, as well as geometric factors, were analyzed and compared against the experimental results of \cite{zhao_dependence_2005}. The work establishes an excellent detailed chemistry model for both surface chemistry as well as electrochemistry with elementary mass action kinetics, which we adopt for our simulations. 
In higher dimensions, \cite{li_simulation_2003} has computed steady state $\rm 2D$ axisymmetric flow in tubular SOFCs with energy and averaged mass transfer as well as simplified chemistry. 
Some $\rm 3D$ simulations of SOFCs have also been performed by various authors. For example, the steady state simulations by \cite{cordiner_analysis_2007} with approximate chemistry, mass averaged diffusion, fluid momentum and energy equations to compute the composition in gas channels as well as porous electrodes. 
\cite{danilov_cfd-based_2009} presented a $\rm 3D$ CFD model for a planar SOFC with internal reforming for studying the influence of various factors on flow field design and kinetics of chemical and electrochemical reactions. 
Electrochemical reactions were computed at the catalyst-electrolyte interface, described by the approximate Butler–Volmer equations for current. 
Possibly representing state of the art is the $\rm 3D$ non-isothermal model for anode‐supported planar SOFC in \cite{li_threedimensional_2019}. The mass, momentum, species, ion, electric, and heat transport equations were solved simultaneously
for co‐flow and counter‐flow arrangements. The Butler‐Volmer chemistry was approximated by Tafel kinetics and the study took the effects of molecular diffusion and Knudsen diffusion into account.

In the works mentioned so far, the microstructure of the electrodes was not resolved but approximated by averaged macroscopic properties like porosity and tortuosity. In \cite{shikazono_numerical_2010}, a $\rm 3D$ numerical simulation of the SOFC was conducted in a resolved microstructure which had been reconstructed by dual-beam focused ion beam–scanning electron microscopy. Gas, ion, and electron transport equations were solved by the LBM in conjunction with electrochemical reactions at the resolved three-phase boundary. The predicted anode overpotential agreed well with the experimental data, although the electrochemistry was calculated with fitted data from the patterned anode experiments of \cite{boer_sofc_1998}. This non exhaustive list of studies reveal an interesting trend. As the physical dimension and the scope of science being replicated expands, the models get simpler. This is expected since not only the computational cost but also the complexity associated with coupling the diverse multi-physics models becomes a challenge. From a numerical standpoint, more differential equations also mean more errors due to the spatial and temporal discretization. The LBM does away with spatial discretization of fluxes and is efficiently scalable due to its simple nearest neighbour interaction, making it a good candidate for complex multi-physics simulations.   

The LBM, which facilitates efficient transient simulations around complex geometries without the hassle of mesh generation, has been successful in modeling multicomponent diffusion with coupled fluid dynamics, species transport, energy and detailed chemistry with full thermodynamic consistency \citep{sawant_consistent_2022,sawant_consistent_2021}. Diffusion modeling with multicomponent pairwise interaction such as with the Stefan-Maxwell model (SMM) or the Dusty Gas Model (DGM) has been shown to be a better approach than modeling diffuson with mass averaging as is done in the Fick's model, especially for experimental validation of fuel cell models \citep{yakabe_evaluation_2000,suwanwarangkul_performance_2003,tseronis_modelling_2008}. However, due to the complexity and cost of the multicomponent diffusion models, Fick's law is mostly preferred in both steady-state \citep{kim_polarization_1999,ferguson_three-dimensional_1996} as well as dynamic simulations \citep{bhattacharyya_dynamic_2009,qi_dynamic_2006}. We have created and validated a transient $\rm 3D$ multicomponent flow solver in \cite{sawant_consistent_2021} which exploits the kinetic nature of the lattice Boltzmann method (LBM) to efficiently model Stefan-Maxwell diffusion, correctly capturing transient reverse diffusion \citep{krishna_review_1997,toor_diffusion_1957}. The model has been extended to reactive flows in \cite{sawant_consistent_2022} and validated with $\rm 3D$ transient simulations in microcombustors with elementary mass action kinetics for hydrogen-air combustion. In SOFC simulations, semi-empirical Butler-Volmer kinetics are often used as a simpler and less demanding alternative to computing the elementary reactions occuring at the triple phase boundary (TPB) \citep{hecht_methane_2005,decaluwe_importance_2008,zhu_modeling_2005}. Considering the current state of lattice Boltzmann modeling, there is an opportunity to create a LBM model to simulate SOFCs with both accurate diffusion as well as detailed chemistry. 
Such a model has the potential to predict concentration polarization due to its sophisticated diffusion model as well as predict activation polarization due to the inclusion of detailed reactions \citep{bhattacharyya_review_2009}. The transient nature of the simulations could reveal the behaviour of the SOFCs under dynamic loading as well as provide an opportunity to study and optimize the startup behaviour of SOFCs \citep{bhattacharyya_review_2009}. 
Since the LBM can already simulate microreactor combustion, reforming reactions in the electrodes and in the flow channels can also be readily accommodated \citep{cordiner_analysis_2007,janardhanan_cfd_2006}. Such a model can also be especially useful for accurate prediction of cell performance at high current densities when multiple species and thermal gradients result from internal reforming \citep{iora_comparison_2005,rostrupnielsen_internal_1995,achenbach_three-dimensional_1994}. In this paper, with this big picture in mind, we have taken the first step of creating a model capable enough of predicting polarization curves of porous SOFCs electrodes through simulation.

The motivation for this work and the justification for adopting the LBM being well stated, we proceed to a more formal introduction. Within the context of computational fluid dynamics (CFD), the lattice Boltzmann method (LBM) \citep{higuera_lattice_1989,higuera_boltzmann_1989} solves a discrete realization of the Boltzmann transport equation \citep{grad_kinetic_1949} at the mesoscale such that the Navier Stokes equations are recovered at the macroscale \citep{chapman_mathematical_1970}. The LBM can simulate a variety of flows including but not limited to   transitional flows, flows in complex moving geometries, compressible flows, multiphase flows, multicomponent flows, rarefied gases, nanoflows etc. \citep{succi_lattice_2018,kruger_lattice_2017,sharma_current_2020}. 
In the preceding works, a model for reactive multicomponent flows has been developed in which a $M$ component fluid is represented by $M$ kinetic equations that model their Stefan-Maxwell interaction, a kinetic equation which models the total mass and mean momentum of the whole fluid and a kinetic equation which models the total energy of all components that make up the multicomponent fluid. In this work, we propose a set of $M$ kinetic equations for the Dusty Gas Model (DGM) \citep{krishna_review_1997} that model Knudsen diffusion as well as Stefan-Maxwell diffusion in a porous medium. The corresponding mean field kinetic equations for the momentum and the energy are formulated to model the representative elementary volume (REV) scale homogenized Navier Stokes equations \citep{whitaker_method_1999}. Together, the $M+2$ system of equations model multicomponent reactive flow in porous media, without resolving the subgrid scale microstructure of the solid matrix. The model is first validated by replicating experimental results for ternary counterdiffusion in capillay tubes over a wide range of rarefied pressures. Next, since only the bulk fluid phase species are modeled by the kinetic equations, source terms are developed to correctly model the interchange of mass and energy between the bulk fluid phase species and the surface phase adsorped species. The detailed chemical mechanism from \cite{decaluwe_importance_2008} is used to introduce the adsoption/desorption as well as electrochemical redox reactions into the model though the open source Cantera \citep{goodwin_cantera_2018} package. The resulting program is used to compute reactive flow in the porous electrodes of a solid oxide fuel cell (SOFC). The current density obtained at different porosities and potentials is validated against the polarization curves from experiments of \cite{zhao_dependence_2005}.    


The paper is structured as follows: We begin with a recap of the nomenclature and the kinetic system for the bulk fluid phase species in section \ref{sec:stefanMaxwell}. 
This section presents the dusty gas model (DGM) discrete lattice Boltzmann equations for the species and their implementation on the standard lattice.  
Next, we turn our attention to describing the representative elementary volume (REV) scale mean field approach for modeling the momentum and energy of the reactive mixture flowing through porous media in section \ref{sec:lbMixtures}. 
Here, we also discuss the realization on standard lattice with the two-population approach. The section closes with a presentation of the resultant macroscopic homogenized Navier-Stokes equations for porous media in the continuum limit. Later, in section \ref{sec:electrodeModel}, we introduce the equations for charge transport as well as the source terms which are necessary to be introduced into the kinetic equations in order to correctly account for the mass and energy changes due to heterogeneous chemical reactions with the adsorbed species. In section \ref{sec:dgm}, the model is used to simulate ternary diffusion in a capillary tube in order to validate the dusty gas model formulation and implementation. Following the check of the diffusion sub-component, we discuss the simulation of a SOFC membrane electrode assembly with the resultant model and validate it against experiments in section \ref{sec:sofc}. We sum up the contributions to the LBM and to the area of fuel cell simulation of the preceding sections and discuss the future possibilities in section \ref{sec:conclusion}. In Appendix \ref{sec:ceNavierStokes} at the end of the paper, we present the Chapman-Enskog analysis which maps the mesoscale kinetic equations to the macroscale homogenised Navier Stokes equations.
\section{Lattice Boltzmann model for the species}
\label{sec:stefanMaxwell}
\subsection{Kinetic equations for the species}
For completeness, we begin with a recap of the nomenclature established in \cite{sawant_consistent_2021} and \cite{sawant_consistent_2022}. The composition of a reactive mixture of $M$ components is described by the fluid phase species densities $\rho_a$, $a=1,\dots, M$, while the mixture density is,
\begin{align}
\rho=\sum_{a=1}^{M}\rho_a.
\label{eq:rhoSpec}
\end{align}
 The rate of change of species densities due to the {homogeneous gas phase} reactions, $\dot \rho_a^{\rm r}$, satisfies mass conservation,
 \begin{equation}
 \sum_{a=1}^{M}\dot \rho_a^{\rm r} = 0.
 \label{eq:sumRhoDot}
 \end{equation}
With the mass fraction,
$Y_a={\rho_a}/{\rho}$ and $m_a$ being the molar mass of the component $a$, the  molar mass of the mixture $m$ is given by
$
{m}^{-1}=\sum_{a=1}^M Y_a/m_a. 
$
The ideal gas equation of state (EoS) provides a relation between the pressure $P$, the temperature $T$ and the composition,
\begin{equation}
P=\rho R T,
\label{eqn:eosIdealGas}
\end{equation}
where $R={R_U}/{m}$ is the specific gas constant of the mixture and $R_U$ is the universal gas constant.
The pressure of an individual component is related to the pressure of the mixture through Dalton's law of partial pressures, 
$P_a=X_a P$, where the mole fraction is $X_a={m} Y_a /{m_a}$. The partial pressure takes the form $P_a=\rho_a R_a T$,
where $R_a={R_U}/{m_a}$ is the specific gas constant of the component.

{In a representative elementary volume (REV) of the porous media, the ratio of the fluid volume to the total volume is represented by the porosity $\phi$. In this work, the porosity is homogeneous in space and time invariant}. In the kinetic representation, each component is described by a set of populations $f_{ai}$ corresponding to the discrete velocities $\bm{c}_i$, $i=0,\dots, Q-1$. The 
species densities $\rho_a$ and the partial momenta $\rho_a \bm{u}_a$ are defined accordingly,
\begin{align}
&	\rho_a {\phi} = \sum_{i=0}^{Q-1}f_{ai},\label{eq:density1}\\
&	\rho_a {\bm{u}_a}= \sum_{i=0}^{Q-1} f_{ai}\bm{c}_i,
	\label{eq:momentum1}
\end{align}
while partial momenta sum up to the mixture momentum, 
\begin{align}
	\rho\bm{u}=\sum_{a=1}^M\rho_a\bm{u}_a.
	\label{eq:totmomSpec}
	\end{align}
{In this work, the choice of placing porosity parameter $\phi$ into various moments was made such that the homogenized Navier Stokes equations \citep{whitaker_method_1999} are correctly recovered at the macro scale. The reader is cautioned that although other choices, for example, the zeroth moment (\ref{eq:density1}) being $\rho$ and the first moment (\ref{eq:momentum1}) being $\rho u_\alpha / \phi$ recovers the correct continuity equation, it does not lead to the expected momentum equation as per the literature on Navier--Stokes equations for homogenized porous media. Although we do not emphasise this repeatedly throughout the paper, the porosity parameter $\phi$ has been carefully placed along with the thermodynamic parameters and moments such as the pressure, enthalpy, flux of enthalpy etc., to recover the homogenized Navier--Stokes equations in the hydrodynamic limit without resorting to ad-hoc forcing terms in the kinetic equations. The form of these moments have been arrived at by repeatedly performing the Chapman--Enskog analysis on the kinetic equations using intuitive guesses for the form of equilibrium moments.}

Starting with kinetic equations in \cite{sawant_consistent_2022} for reactive species with Stefan-Maxwell diffusion, we modify the equations to include Knudsen diffusion. The kinetic equations for the species are written as,
\begin{align}
\partial_t f_{ai} + \bm{c}_{i}\cdot \nabla f_{ai} = \sum_{b\ne a}^M \frac{PX_aX_b}{\mathcal{D}_{ab}} \left[ \left( \frac{f_{ai}^{\rm eq}-f_{ai}}{\rho_a} \right) - \left( \frac{f_{bi}^{\rm eq}-f^*_{bi}}{\rho_b} \right) \right] 
{- \frac{P X_a}{ \mathcal{D}_a^{\rm k}} \left(  \frac{f^*_{ai}-f_{ai}^{\rm k}}{\rho_a} \right)} + \dot f_{ai}^{\rm r}, 
\label{eqn:stefanMaxwell} 
\end{align}
%
%
%
where  $\mathcal{D}_{ab}$ are Stefan--Maxwell binary diffusion coefficients, {$\mathcal{D}_{a}^{\rm k}$ are the Knudsen diffusion coefficients}, while the reaction source population $\dot f_{ai}^{\rm r}$ satisfies the following conditions:
\begin{align}
& \sum_{i=0}^{Q-1} \dot f_{ai}^{\rm r}= \dot \rho_a^{\rm r}, \label{eq:Dotfr}\\
&\sum_{i=0}^{Q-1} \dot f_{ai}^{\rm r}\bm{c}_i =  \dot \rho_a^{\rm r} \frac{\bm{u}}{{\phi}}.
 \label{eq:m1fDotc}
\end{align}
%
%
We now proceed with specifying the  equilibrium $f_{ai}^{\rm eq}$, the quasi-equilibrium populations $f^*_{ai}$ and $f^k_{ai}$, and the reaction source populations. 
\subsection{Standard lattice and product-form} \label{sec:StandardLattice}
All the kinetic models including (\ref{eqn:stefanMaxwell}) are realized on the standard discrete velocity set $D3Q27$, where $D=3$ stands for three dimensions and $Q=27$ is the number of discrete velocities,
\begin{equation}\label{eq:d3q27vel}
	\bm{c}_i=(c_{ix},c_{iy},c_{iz}),\ c_{i\alpha}\in\{-1,0,1\}.
\end{equation}
In order to specify the equilibrium $f_{ai}^{\rm eq}$, the quasi-equilibriums $f^*_{ai}$ and $f^k_{ai}$, and the reaction source term $\dot f_{ai}^{\rm r}$ in (\ref{eqn:stefanMaxwell}), we first define a triplet of functions in two variables, $\xi_{\alpha}$ and $\PP_{\alpha\alpha}$, 
\begin{align}
&	\Psi_{0}(\xi_{\alpha},\PP_{\alpha\alpha}) = 1 - \PP_{\alpha\alpha}, 
	\label{eqn:phi0}
	\\
&	\Psi_{1}(\xi_{\alpha},\PP_{\alpha\alpha}) = \frac{\xi_{\alpha} + \PP_{\alpha\alpha}}{2},
	\label{eqn:phiPlus}
	\\
&	\Psi_{-1}(\xi_{\alpha},\PP_{\alpha\alpha}) = \frac{-\xi_{\alpha} + \PP_{\alpha\alpha}}{2},
	\label{eqn:phis}
\end{align}
%
and consider a product-form associated with the discrete velocities $\bm{c}_i$ (\ref{eq:d3q27vel}),
\begin{equation}\label{eq:prod}
	\Psi_i= \Psi_{c_{ix}}(\xi_x,\PP_{xx}) \Psi_{c_{iy}}(\xi_y,\PP_{yy}) \Psi_{c_{iz}}(\xi_z,\PP_{zz}).
\end{equation}
All pertinent populations are determined by specifying the parameters 
$\xi_\alpha$ and $\PP_{\alpha\alpha}$ in the product-form (\ref{eq:prod}). 
The equilibrium  and the quasi-equilibrium populations are,
%
%
\begin{align}
	f_{ai}^{\rm eq}
	&= {\phi} \rho_a\Psi_{c_{ix}}\left(\frac{u_x}{{\phi}},\frac{u_x^2}{{\phi^2}}+R_aT\right) \Psi_{c_{iy}}\left(\frac{u_y}{{\phi}},\frac{u_y^2}{{\phi^2}}+R_aT\right) \Psi_{c_{iz}}\left(\frac{u_z}{{\phi}},\frac{u_z^2}{{\phi^2}}+R_aT\right),
\label{eq:27eq}	\\
	f_{ai}^{*}
	&= {\phi} \rho_a\Psi_{c_{ix}}\left(\frac{u_{ax}}{{\phi}},\frac{u_{ax}^2}{{{\phi^2}}}+R_aT\right)
	\Psi_{c_{iy}}\left(\frac{u_{ay}}{{\phi}},\frac{u_{ay}^2}{{{\phi^2}}}+R_aT\right) 
	\Psi_{c_{iz}}\left(\frac{u_{az}}{{\phi}},\frac{u_{az}^2}{{{\phi^2}}}+R_aT\right).
\label{eq:27qeq}
\end{align}
The product-form (\ref{eq:prod}) together with the equilibrium parameters 
are used to specify the reaction terms,
\begin{align}
    	\dot f_{ai}^{\rm r}
    	&= \dot \rho_a^{\rm r}  \Psi_{c_{ix}}\left(\frac{u_{x}}{{\phi}},\frac{u_{x}^2}{{{\phi^2}}}+R_aT\right)
	\Psi_{c_{iy}}\left(\frac{u_{y}}{{\phi}},\frac{u_{y}^2}{{\phi^2}}+R_aT\right) 
	\Psi_{c_{iz}}\left(\frac{u_{z}}{{\phi}},\frac{u_{z}^2}{{\phi^2}}+R_aT\right),
\label{eq:fsource} 
\end{align}
{ while the quasi-equilibrium populations responsible for enabling Knudsen diffusion are  specified as,
\begin{align}
    	f_{ai}^{\rm k}
    	&= \phi \rho_a  \Psi_{c_{ix}}\left(0,R_aT\right)
	\Psi_{c_{iy}}\left(0,R_aT\right) 
	\Psi_{c_{iz}}\left(0,R_aT\right).
\label{eq:fKnudsen}
\end{align} }
{The populations $f_{ai}^{\rm k}$ have a zero velocity while the populations $f^*_{ai}$ from which they are subtracted in the kinetic equation (\ref{eqn:stefanMaxwell}) have all of the component velocity. Intuitively, the populations $f_{ai}^{\rm k}$ can be thought of as representing the stationary component, consistent with the idea of the dusty gas model \citep{mason_statistical-mechanical_1990}. The term $(f^*_{ai}-f_{ai}^k)$ represents a retardation proportional velocity of component $a$ due to its interaction with the stationary component.}

Along the lines of \cite{sawant_consistent_2021}, analysis of the hydrodynamic limit of the kinetic model (\ref{eqn:stefanMaxwell}) leads to the following: The balance equations for the densities of the species in the presence of the source term are found as follows,
\begin{align}
	\partial_t  \phi \rho_a + \nabla\cdot(\rho_a \bm{u}_a) = \dot \rho_a^{\rm r},
	\label{eq:dtrhoa}
\end{align}
where the component velocities, $\bm{u}_a$ satisfy the Stefan--Maxwell constitutive relation with {Knudsen diffusion} \citep{mason_statistical-mechanical_1990,krishna_review_1997,kee_chemically_2003},
\begin{equation}
	P\nabla X_a+(X_a-Y_a)\nabla P=\sum_{b\ne a}^M \frac{PX_aX_b}{ {\phi} \mathcal{D}_{ab}} \left(\bm{u}_{b} - \bm{u}_a \right) { - \frac{P X_a}{\phi \mathcal{D}_{a}^{\rm k}} \bm{u}_{a}}. 
	\label{eq:constit2}
\end{equation}
%
Summarizing, the kinetic model (\ref{eqn:stefanMaxwell}) recovers the dusty gas model with a provision for modeling composition changes due to chemical reactions.

\subsection{Lattice Boltzmann equation for the species}

The kinetic model (\ref{eqn:stefanMaxwell}) is transformed into a lattice Boltzmann equation by following a process similar to the one for Stefan-Maxwell diffusion case detailed in \cite{sawant_consistent_2021,sawant_consistent_2022}. 
Upon integration of \eqref{eqn:stefanMaxwell} along the characteristics and application of the trapezoidal rule to all relaxation terms on the right hand side except for the reaction term, we arrive
at a fully discrete lattice Boltzmann equation for the species,
\begin{equation}
	f_{ai}(\bm{x}+\bm{c}_i \delta t, t+ \delta t)  = f_{ai}(\bm{x},t)+ 2 \beta_a [f_{ai}^{\rm eq}(\bm{x},t) - f_{ai}(\bm{x},t)]
	+ \delta t (\beta_a-1) F_{ai}(\bm{x}, t) +  R_{ai}^{\rm r}.
	\label{eqn:finalNumericalEquationsReactive}
\end{equation}
Here $\delta t$ is the lattice time step, the equilibrium populations are provided by Eq.\ (\ref{eq:27eq}), while the relaxation parameters $\beta_a\in[0,1]$ are,
\begin{equation}
	\beta_a=\frac{\delta t}{2 \tau_a + \delta t}.
	\label{eqn:betaa}
\end{equation}
Their relation to the Stefan--Maxwell binary diffusion coefficients is found as follows:
Introducing characteristic times,
\begin{equation}
	\tau_{ab}=\frac{mR_UT}{\mathcal{D}_{ab}m_a m_b},
 \label{eqn:tauab}
\end{equation} 
the relaxation times $\tau_a$ in (\ref{eqn:betaa}) are defined as, 
\begin{equation}
	\frac{1}{\tau_a} = \sum_{b\ne a}^M \frac{Y_b}{\tau_{ab}}.
	\label{eqn:taua}
\end{equation}
In (\ref{eqn:finalNumericalEquationsReactive}), the quasi-equilibrium relaxation term $F_{ai}$,
is spelled out as follows,
\begin{equation}
F_{ai} = Y_a \sum_{b\ne a}^M \frac{1}{\tau_{ab}}  \left( f_{bi}^{\rm eq}-f_{bi}^* \right) { + \frac{1}{\tau_{a}^{\rm k}}  \left( f^*_{ai}-f_{ai}^{\rm k} \right)} 
\label{eqn:fStar}
\end{equation}
%
%
{The relaxation times corresponding to Knudsen diffusion $\tau_a^{\rm k}$ in (\ref{eqn:fStar}) are defined as, 
\begin{equation}
	\tau_{a}^{\rm k}=\frac{R_U T}{\mathcal{D}_{a}^{\rm k} m_a }.
	\label{eqn:tauak}
\end{equation}}
The quasi-equilibrium populations $f_{bi}^*$ in (\ref{eqn:fStar}) are defined by the product-form (\ref{eq:27qeq}), subject to the following parameterization,
\begin{align}
	&\xi_\alpha=u_{\alpha}+V_{b\alpha},\\ &\PP_{\alpha \alpha}=R_bT+\left(u_{\alpha}+V_{b\alpha}\right)^2,
	\label{eq:xiPstar2}
\end{align}
where the second-order accurate diffusion velocity $\bm{V}_b$ is the result of the lattice Boltzmann discretization of the kinetic equation and is found by solving the $M\times M$ linear algebraic system for each spatial component,
\begin{align}
\left( 1+ \frac{\delta t}{2 \tau_a} { + \frac{\delta t}{2 \tau_a^{\rm k}} } \right)  {\bm{V}_{a}} - \frac{\delta t}{2} \sum_{b\ne a}^{M} \frac{1}{\tau_{ab}} Y_b  {\bm{V}_{b}} 
=\bm{u}_{a}-\bm{u}.
\label{eqn:transform1}
\end{align}
%
Equation (\ref{eqn:transform1}) can be written in a more compact form as,
\begin{align}
\left( 1+ \frac{\delta t}{2 \tau_a} { + \frac{\delta t}{2 \tau_a^{\rm k}} } \right)  {\bm{V}_{a}} - \frac{\delta t}{2} \sum_{b}^{M}  \left( \frac{1-\delta_{ab}}{\tau_{ab}} 
\right) 
Y_b  {\bm{V}_{b}}    =\bm{u}_{a}-\bm{u}.
\label{eqn:transform1Alternate}
\end{align}
%
%
The system (\ref{eqn:transform1}) has been altered by the inclusion of Knudsen diffusion, therefore, it is different from the form which was proposed in the earlier works, e.g. \cite{sawant_consistent_2022}. 
In our realization, we solve (\ref{eqn:transform1}) with the Householder QR decomposition method from the Eigen library \citep{guennebaud_eigen_2010}. 

 Finally, the reaction term in (\ref{eqn:finalNumericalEquationsReactive}) is represented by an integral over the characteristics,
\begin{align}
	{R_{ai}^{\rm r}}=\delta t\int_{0}^{1}\dot f_{ai}^{\rm r}(\bm{x}+\bm{c}_i s\delta t,t+s\delta t )ds.
	\label{eq:R1}
\end{align}
Taking into account the structure of the reaction term (\ref{eq:fsource}), we use a simple explicit approximation for the implicit term (\ref{eq:R1}),
\begin{align}
	{R_{ai}^{\rm r}}\approx \dot f_{ai}^{\rm r}(\bm{x},t) \delta t.
	\label{eq:R2}
\end{align}
Reaction rates $\dot \rho_a^{\rm r}$ are obtained from the open source chemical kinetics package Cantera \citep{goodwin_cantera_2018} as a function of mixture internal energy $U$ and composition, $\dot \rho_a^{\rm r}=\dot \rho_a^{\rm r}(U,\rho_1,\dots,\rho_M)$. 

Summarizing, the lattice Boltzmann system (\ref{eqn:finalNumericalEquationsReactive}) delivers the extension of the species dynamics to the dusty gas model in reactive mixtures. We now proceed with setting up the lattice Boltzmann equations for the mixture momentum and energy.
\section{Lattice Boltzmann model of mixture momentum and energy}
\label{sec:lbMixtures}
%
%
\subsection{Double-population lattice Boltzmann equation}
\label{sec:energy}
The mass-based specific internal energy ${U}_{a}$ and enthalpy ${H}_{a}$ of the species are,
\begin{align}
	{U}_{a}&=U^0_a+\int_{T_0}^T{C}_{a,v}(T')dT',
	\label{eq:specUa}\\
	{H}_{a}&=H^0_a+\int_{T_0}^T{C}_{a,p}(T')dT',
	\label{eq:specHa}
\end{align}
where $U^0_a$ and  $H^0_a$ are the energy and the enthalpy of formation at the reference temperature $T_0$, respectively, 
while $C_{a,v}$ and $C_{a,p}$ are specific heats at constant volume and at constant pressure, satisfying the Mayer relation, ${C}_{a,p}-{C}_{a,v}=R_{a}$.
Consequently, the internal energy $\rho U$ and enthalpy $\rho H$ of the mixture are defined as,
\begin{align}
	\rho U=\sum_{a=1}^M\rho_a U_a,
		\label{eq:U}\\
	 \rho H=\sum_{a=1}^M\rho_a H_a.
	\label{eq:H}
\end{align}
We follow a two-population approach \citep{he_novel_1998,guo_thermal_2007,karlin_consistent_2013,frapolli_entropic_2018}. One set of populations ($f$-populations) is used to represent the density and the momentum of the mixture.
Below, we refer to the $f$-populations as the momentum lattice.  The locally conserved fields are the volume fraction of density and the momentum of the mixture,
\begin{align}
&\sum_{i=0}^{Q-1} f_i = \sum_{i=0}^{Q-1} f_i^{\rm eq} = {\phi} \rho,\label{eqn:fdensity}\\
&\sum_{i=0}^{Q-1} f_i \bm{c}_{i} = \sum_{i=0}^{Q-1} f_i^{\rm eq} \bm{c}_{i} =  \rho {\bm{u}}.
\label{eqn:f1momMomentum} 
\end{align}
Another set of populations ($g$-populations), or the energy lattice, is used to represent the local conservation of the volume fraction of total energy of the mixture,
\begin{align}
&\sum_{i=0}^{Q-1} g_i  = \sum_{i=0}^{Q-1} g_i^{\rm eq} =  \phi \rho E,  \label{eqn:g0momTotalEnergy} \\
&  \phi \rho E= {\phi} \rho \left( U + {\frac{u^2}{2 {\phi^2}}} \right).
\label{eq:totalE}
\end{align}
The species kinetic equations are coupled with the kinetic equations for the mixture through the dependence of mixture internal energy (\ref{eq:U}) on the composition. From (\ref{eq:specUa}), (\ref{eq:U}) and (\ref{eq:totalE}), the temperature is evaluated by solving the  integral equation,
\begin{equation}
\label{eq:temperature}
  \sum_{a=1}^MY_a\left[U_a^0 + \int_{T_0}^T {C}_{a,v}(T')dT'\right]=E-\frac{u^2}{2 {\phi}}.
\end{equation}
The temperature evaluated by solving (\ref{eq:temperature}) enters the species lattice Boltzmann system through the pressure (\ref{eqn:eosIdealGas}),   forming a two-way coupling.

The lattice Boltzmann equations for the momentum and for the energy lattice are patterned from the single-component developments \citep{saadat_extended_2021} and are realized on the $D3Q27$ discrete velocity set.
The mixture lattice Boltzmann equations are written,
\begin{align}
f_i(\bm{x}+\bm{c}_i \delta t,t+\delta t)- f_i(\bm{x},t)&=  \omega (f_i^{\rm ex} -f_i),  \label{eqn:f} 
\\
g_i(\bm{x}+\bm{c}_i \delta t,t+ \delta t) - g_i(\bm{x},t)&=  \omega_1 (g_i^{\rm eq} -g_i) + (\omega - \omega_1) (g_i^{*} -g_i),
 \label{eqn:g}
\end{align}
where relaxation parameters $\omega$ and $\omega_1$ are related to the mixture viscosity and thermal conductivity, and we proceed with specifying the pertinent populations in (\ref{eqn:f}) and (\ref{eqn:g}).

\subsection{Extended equilibrium for the momentum lattice}
\label{sec:ExtendedEquilibriumMmomentum}

The extended equilibrium populations $f_i^{\rm ex}$ in (\ref{eqn:f}) are specified by the product-form (\ref{eq:prod}), 
with the parameters identified  as ${\xi}_{\alpha}={{u}_{\alpha}^{\rm ex}/\phi}$ and $\PP_{\alpha \alpha}=\PP_{\alpha \alpha}^{\rm ex}$,
\begin{equation}
f_{i}^{\rm ex}= \phi \rho 
\Psi_{c_{ix}}\left({\frac{u_{x}^{\rm ex}}{\phi}},\PP_{xx}^{\rm ex}\right)
\Psi_{c_{iy}}\left({\frac{u_{y}^{\rm ex}}{\phi}},\PP_{yy}^{\rm ex}\right) 
\Psi_{c_{iz}}\left({\frac{u_{z}^{\rm ex}}{\phi}},\PP_{zz}^{\rm ex}\right),
\label{eq:feqmix}
\end{equation}
where the extended parameter $\PP_{\alpha \alpha}^{\rm ex}$ reads,
\begin{align}
	\PP_{\alpha \alpha}^{\rm ex}& = \PP_{\alpha \alpha}^{\rm eq}
 + \left[ - \varsigma + \mu \left( \frac{2}{D} - \frac{R}{C_v} \right) \right] \frac{(\partial_\beta u_\beta)}{\phi \rho}
 + \delta t\left( \frac{2-\omega}{2 \phi\rho\omega}\right)
	\partial_\alpha \left(\rho u_\alpha (1 - 3 R T) - {\frac{\rho u_\alpha^3}{\phi^2}}\right).
	\label{eqn:Pex}
\end{align}
{and the extended velocity $u_{\alpha}^{\rm ex}$ which models the effect of permeability through the force density due to Knudsen diffusion $\bm{\mathcal{F}}^{\rm k}$ reads,
\begin{align}
u_{\alpha}^{\rm ex}=u_{\alpha} \left( 1 + \frac{\delta t}{\omega}  \frac{1}{\rho}  \mathcal{F}_\alpha^{\rm k} \right), 
\end{align}}
while $\PP_{\alpha \alpha}^{\rm eq}$ is, 
\begin{align}
	\PP_{\alpha \alpha}^{\rm eq}& =  RT+\frac{u_\alpha^2}{{\phi^2}}.
	\label{eqn:Peq}
\end{align}
%
%
As is visible from the second last term in (\ref{eqn:stefanMaxwell}), the action of the Knudsen diffusion is to introduce a forcing on the species which would not vanish when the momentum represented by equation (\ref{eqn:stefanMaxwell}) is summed over all the components. Therefore, the term $\bm{\mathcal{F}}^{\rm k}$ has been used to introduce a correction to the hydrodynamic flux. It is computed as,
\begin{align} \label{eqn:netKnudsenForce}
    \bm{\mathcal{F}}^{\rm k}=-\sum_{b=1}^M \frac{\phi P X_b}{\mathcal{D}_{b}^{\rm k}} \frac{\bm{u}_{b}}{\phi}
\end{align}
   
The effect of extension, featured by the third term in (\ref{eqn:Pex}), is to correct for the incomplete Galilean invariance of the standard $D3Q27$ velocity set (\ref{eq:d3q27vel}). The second term in (\ref{eqn:Pex}) is necessary to impose the correct bulk viscosity $\varsigma$ \citep{sawant_consistent_2022}. 
The equilibrium pressure tensor $P_{\alpha \beta}^{\rm eq}$ is given by,
\begin{align} \label{eq:Peq}
    \sum_{i=0}^{Q-1} f_i^{\rm eq} c_{i\alpha} c_{i\beta} = P_{\alpha \beta}^{\rm eq} =  \phi \rho \frac{u_\alpha}{ \phi} \frac{u_\beta}{ \phi} +  \phi P \delta_{\alpha \beta}.
\end{align}
%
%

\subsection{Equilibrium and quasi-equilibrium of the energy lattice}
\label{sec:ExtendedEquilibriumEnergy}

For the energy lattice, the corresponding equilibrium and quasi-equilibrium populations in (\ref{eqn:g}) are evaluated along the lines of \cite{saadat_extended_2021} using linear operators $\mathcal{O}_\alpha$, acting on any smooth function $A(\bm{u},T)$ according to a rule,
\begin{equation}
    \mathcal O_\alpha A= 
    RT \frac{\partial A}{\partial u_\alpha} + u_\alpha A.
    \label{eqn:oalpha}
\end{equation}
By substituting the parameters $\xi_\alpha = \mathcal{O}_\alpha$ and $\PP_{\alpha\alpha} = \mathcal{O}_\alpha^2$ into the product form (\ref{eq:prod}), the equilibrium populations $g_i^{\rm eq}$ are compactly written using the energy $E$ as the generating function,
\begin{align}
    g_i^{\rm eq} = \phi \rho \Psi_{c_{ix}}(\mathcal{O}_x,\mathcal{O}_x^2) \Psi_{c_{iy}}(\mathcal{O}_y,\mathcal{O}_y^2) \Psi_{c_{iz}}(\mathcal{O}_z,\mathcal{O}_z^2)E. \label{eq:geq_i}
\end{align}
A direct computation of the equilibrium (\ref{eq:geq_i}) satisfies the necessary conditions to recover the mixture energy equation, namely, the equilibrium energy flux $\bm{q}^{\rm eq}$ and the flux thereof $\bm{R}^{\rm eq}$, 
 \begin{align}
 &\bm{q}^{\rm eq}= \sum_{i=0}^{Q-1}  g_i^{\rm eq} \bm{c}_{i} =  \left(H+\frac{u^2}{2 {\phi^2}}\right)\rho\bm{u},
 \label{eqn:geq1mom} 
 \\
 &\bm{R}^{\rm eq}=\sum_{i=0}^{Q-1} g_i^{\rm eq} \bm{c}_i\otimes\bm{c}_i =
     \left(H+\frac{u^2}{2 {\phi^2}}\right) \bm{P}^{\rm eq} + \frac{P}{ \phi}\bm{u}\otimes\bm{u},
 \label{eqn:geq2mom}
 \end{align}
where $H$ is the specific mixture enthalpy (\ref{eq:H}). 
The quasi-equilibrium populations $g_i^*$ 
differs from the equilibrium $g_i^{\rm eq}$ by the energy flux only  \citep{karlin_consistent_2013,sawant_consistent_2021,saadat_extended_2021},
	\begin{align}
		g_{i}^*= \left\{\begin{aligned}
			& g_{i}^{\rm eq}+\frac{1}{2}\bm{c}_i\cdot\left(\bm{q}^*-\bm{q}^{\rm eq}\right), &\text{ if } c_i^2=1, & \\ 
			&g_i^{\rm eq}, & \text{otherwise}.&\\
		\end{aligned}\right.
	\label{eq:gstar}	
	\end{align}
where $\bm{q}^*$ is a specified quasi-equilibrium energy flux,
\begin{align}
	\bm{q}^{*} &=\sum_{i=0}^{Q-1} g_i^{*} \bm{c}_{i} =  \bm{q} - \frac{\bm{u}}{ \phi} \cdot (\bm{P} - \bm{P}^{\rm eq}) +\bm{q}^{\rm diff}+\bm{q}^{\rm corr}+\bm{q}^{\rm ex}.
	\label{eq:gstareq1mom}
\end{align}
The two first terms in (\ref{eq:gstareq1mom}) maintain a variable Prandtl number and include the energy flux $\bm{q}$ and the pressure tensor $\bm{P}$,
\begin{align}
&	\bm{q}=\sum_{i=0}^{Q-1} g_i \bm{c}_{i},\\
&	\bm{P}=\sum_{i=0}^{Q-1} f_i \bm{c}_{i}\otimes \bm{c}_{i}.
\end{align}
The interdiffusion energy flux $\bm{q}^{\rm diff}$,
\begin{align} \label{eq:interdiffusion}
\bm{q}^{\rm diff} =\left(\frac{\omega_1}{\omega-\omega_1} \right) \rho\sum_{a=1}^{M}H_aY_a \bm{V}_a,
\end{align}
where the diffusion velocities $\bm{V}_a$ are defined by Eq.\ (\ref{eqn:transform1}), contributes the enthalpy transport due to diffusion, cf.\ \citep{sawant_consistent_2021}.
Moreover, the correction flux $\bm{q}^{\rm corr}$ is required in the two-population approach to the mixtures in order to recover the Fourier law of thermal conduction  \citep{sawant_consistent_2021},
\begin{align}
\bm{q}^{\rm corr}=\frac{1}{2}\left(\frac{\omega_1-2}{\omega_1-\omega}\right) {\delta t}P \phi \sum_{a=1}^M  H_{a} \nabla  Y_a.
\label{eq:corrFourier}
\end{align}
The term $\bm{q}^{\rm ex}$ in the quasi-equilibrium flux  (\ref{eq:gstareq1mom}) is required for consistency with the extended equilibrium (\ref{eq:feqmix}). Components of the vector $\bm{q}^{\rm ex}$ follow the structure of (\ref{eqn:Pex}), 
\begin{align}
	{q}_\alpha^{\rm ex} &= u_\alpha \left( \frac{\omega}{\omega-\omega_1} \right) \left[ - \varsigma + \mu \left( \frac{2}{D} - \frac{R}{C_v} \right) \right] \frac{(\bm{\nabla}\cdot\bm{u})}{\phi}
 - \frac{1}{2}\delta t  \frac{u_\alpha}{\phi}
	\partial_\alpha \left(\rho u_\alpha \left(1 - 3 R T\right) - \frac{\rho u_\alpha^3}{{\phi^2}} \right).
	\label{eqn:qex}    
\end{align}
Spatial derivatives in the correction flux (\ref{eq:corrFourier}) and in the isotropy correction (\ref{eqn:Pex}) and (\ref{eqn:qex}) were implemented using  isotropic lattice operators \citep{thampi_isotropic_2013}.
%

The lattice Boltzmann model for a $M$-component mixture of ideal gas on the standard $D3Q27$ lattice consists of $M$ species lattices where the lattice Boltzmann equation is given by Eq.\ (\ref{eqn:finalNumericalEquationsReactive}), and the momentum and energy lattice Boltzmann equations (\ref{eqn:f}) and (\ref{eqn:g}). In total, the $M+2$ lattice Boltzmann equations are tightly coupled, as has been already specified above: The temperature from the energy lattice is provided to the species lattices through species equilibrium (\ref{eq:27eq}) and quasi-equilibriums (\ref{eq:27qeq}), (\ref{eq:fsource}) and (\ref{eq:fKnudsen}), but also in the Stefan--Maxwell temperature-dependent relaxation rates (\ref{eqn:tauab}) and the Knudsen diffusion temperature-dependent relaxation rates (\ref{eqn:tauak}). 

Looking at the information flowing in the other direction, the net force due to Knudsen diffusion $\bm{\mathcal{F}}^{\rm k}$ computed as (\ref{eqn:netKnudsenForce}) is an input to the to the momentum lattice which relies on the component velocity and composition from the species lattice. The species diffusion velocities are also in input to the quasi-equilibrium population of the energy lattice via the interdiffusion flux (\ref{eq:interdiffusion}). The mass fractions from the species lattices are used to compute the mixture energy (\ref{eq:U}) and enthalpy (\ref{eq:H}) in the equilibrium and the quasi-equilibrium of the momentum and energy lattices. The momentum and the energy lattices remain coupled in the standard way already present in the single-component setting. In this formulation, we use these interconnections between the species, and the momentum and energy lattices, which has been termed as {\it weak coupling} in \cite{sawant_consistent_2022}. It should be mentioned that the other stronger forms of coupling mentioned in \cite{sawant_consistent_2022} cannot be used with this formulation. This is because the stronger forms of coupling mentioned therein eliminate the momentum of one of the species, which would lead to an incorrect $\bm{\mathcal{F}}^{\rm k}$.
\subsection{Mixture mass, momentum and energy equations}
With the equilibrium and quasi-equilibrium populations specified, the hydrodynamic limit of the two-population lattice Boltzmann system (\ref{eqn:f}) and (\ref{eqn:g}) is found by expanding the propagation to second order in the time step $\delta t$ and evaluating the moments of the resulting expansion. The detained analysis is presented in Appendix \ref{sec:ceNavierStokes}.
The continuity equation \citep{whitaker_method_1999}, the momentum equation with penalization \citep{whitaker_method_1999,liu_brinkman_2007,fuchsberger_incorporation_2022} and the energy equations for a reactive multicomponent mixture \citep{kee_chemically_2017,williams_combustion_1985,bird_transport_2007} are, respectively,
\begin{align}
&\partial_t  \phi \rho + \nabla\cdot (\rho \bm{u})=0,
\label{eqn:dtrho}
\\
&\partial_t (\rho\bm{u}) + \frac{1}{ \phi} \nabla\cdot ({\rho\bm{u}\otimes\bm{u} })+ \nabla\cdot \bm{\pi}=\bm{\mathcal{F}}^{\rm k},
\label{eqn:dtu}
\\
&\partial_t ( \phi \rho E)+\nabla\cdot(\rho E\bm{u})+ \nabla\cdot\bm{q}+ \frac{1}{ \phi} \nabla\cdot(\bm{\pi}\cdot\bm{u})=0.
\label{eqn:dtE}
\end{align}
Here, the pressure tensor $\bm{\pi}$ in the momentum equation reads,
\begin{equation}\label{eq:NSmix}
\bm{\pi}=  \phi P\bm{I}
-\mu \left( \nabla\bm{u}  + \nabla\bm{u}^{\dagger}  -\frac{2}{D} (\nabla\cdot\bm{u})\bm{I} \right) 
-\varsigma (\nabla\cdot\bm{u}) \bm{I},
\end{equation}
where the dynamic viscosity $\mu$ is related to the relaxation parameter $\omega$,
\begin{align}
\mu  &= \left( \frac{1}{\omega} - \frac{1}{2}\right) P{\delta t},
\label{eq:mu}
\end{align}

Here $C_v=\sum_{a=1}^M Y_a C_{a,v}$ is the mixture specific heat at constant volume. The heat flux $\bm{q}$ in the energy equation (\ref{eqn:dtE}) reads,
\begin{equation}
\label{eq:qneq}
\bm{q}=-  \phi \lambda\nabla T+\rho\sum_{a=1}^{M}H_aY_a {\bm{V}_a} .
\end{equation}
The first term in (\ref{eq:qneq}) is the Fourier law of thermal conduction in the gas phase \citep{kee_chemically_2017}, with thermal conductivity $\lambda$ related to the relaxation parameter $\omega_1$,
\begin{equation}\label{eq:lambda}
\lambda= \left(\frac{1}{\omega_1} - \frac{1}{2}\right) P C_p{\delta t},
\end{equation}
where $C_p=C_v+R$ is the mixture specific heat at constant pressure.
The second term in (\ref{eq:qneq}) is the interdiffusion energy flux. With the thermal diffusivity $\alpha=\lambda/\rho C_p$ and the kinematic viscosity $\nu=\mu/\rho$, the Prandtl number becomes, 
${\rm Pr} = {\nu}/{\alpha}$.
For this reactive formulation, the local dynamic viscosity $\mu(\bm{x},t)$ and the thermal conductivity $\lambda(\bm{x},t)$ of the mixture is evaluated as a function of the local chemical state by using the chemical kinetics solver Cantera \citep{goodwin_cantera_2018,kee_chemically_2003,wilke_viscosity_1950,mathur_thermal_1967}. 

{
\section{Reactions in porous electrodes}
\label{sec:electrodeModel}

In order to get useful insights from applying the model developed so far to reactive flow in porous electrodes, the model needs to be augmented with some additions pertaining to electrochemical reactions. Some source terms are needed in the kinetic equation for energy (\ref{eqn:g}) to account for ohmic heat, energy lost as electricity and for balancing the energy changes due to interchange of species between the surface phase and the bulk phase.
The kinetic equation for momentum (\ref{eqn:f}) also needs to account for the change of mass caused by adsorption and desorption.

Within a representative elementary volume, let us define $a^{(s)}_k$ as the specific surface area of a reactive surface $k$. The specific surface area is the area available for surface reactions per unit volume of the porous material \citep{kee_chemically_2017}. The product of specific surface area and the surface rate of production of species gives the bulk production rate of the species. For a gas phase species having the density $\rho_a$, it's total mass production rate per unit volume in the gas phase $\dot \rho_a^{\rm (f)}$ is computed as \citep{decaluwe_importance_2008},  
%
\begin{align} \label{eq:totalFluidReactionRate}
    \dot \rho_a^{\rm (f)} = \dot \rho_a^{\rm r} + \sum_{k=1}^\kappa a^{(s)}_k \dot \rho_{a,k} . 
\end{align}
In equation (\ref{eq:totalFluidReactionRate}), $\dot \rho_a^{\rm r}$ is the net mass production rate per unit volume of species $a$ in the fluid phase due to homogeneous reactions and $\dot \rho_{a,k}$ is the net mass production rate of species $a$ due to heterogeneous reactions per unit surface area of the surface $k$,  out of the $\kappa$ number of chemically active surfaces. 
The latter is calculated as,
\begin{align} \label{eq:molarToMassRates}
\dot \rho_{a,k} = m_a \dot M_{a,k}^{\rm (f)}   ,
\end{align}
with $\dot M_{a,k}^{\rm (f)}$ being the molar production rate of the fluid phase species $a$ per unit surface area of the surface material $k$. The molar production rate $\dot M_{a,k}^{\rm (f)}$ is responsible for describing interchange between the aforementioned fluid phase species $a$ and the surface phase species that exist on the surface in an adsorbed state. The composition of the surface species is described by the number of moles per unit area of the adsorbed site $M_{a,k}^{\rm (s)}$ and the constant site density $\Gamma_k$, which is the total capacity of the surface to host adsorbed species. The mole fraction of an adsorbed species is then defined as,
\begin{align} \label{eq:surfaceMoleFractions}
    X_{a,k}^{\rm (s)} = \frac{M_{a,k}^{\rm (s)}}{\Gamma_k}
\end{align}
 Analogous to the surface reactions, the edges formed at the intersection of the surfaces are also capable of hosting chemical reactions. The edges are described by their specific length $l^{(e)}_p$, i.e. the length of the edge $p$ per unit volume of the REV. The net reaction rate of a surface species $a$ on an edge $p$ are then described by the molar production rate per unit length $\dot M_{a,p}^{\rm (e)}$.
 The rate of change of a surface species $a$ residing on the surface $k$ is the non-dimensionalised sum of it's molar production rate on the surface $k$ and it's molar production rate on all the edges $p$ belonging to the surface $k$. Mathematically, the rate of change of mole fraction $\dot X_{a,k}^{\rm (s)}$ is written as \citep{decaluwe_importance_2008},  
 \begin{align} \label{eq:totalSurfaceReactionRate}
      \dot X_{a,k}^{\rm (s)} = \frac{1}{\Gamma_k} \left( \dot M_{a,k}^{\rm (s)} +  \frac{1}{a^{(s)}_k} \sum_{p \in k} l^{(e)}_p \dot M_{a,p}^{\rm (e)} \right)
 \end{align} 
In this work, we solve for the flow through porous electrodes which are made up of spherical microstructures of at most two substances. The anode is made of nickel and yttria stabilized zirconia (YSZ), while the cathode is made up of lanthanum strontium manganite (LSM) and YSZ. The nickel and the LSM forms the electrode phase in the anode and the cathode, respectively. The YSZ forms the electrolyte phase in both the anode and the cathode. The intersection of the micro spheres of the electrode and the electrolyte phase form an edge, which is also referred to as the triple phase boundary (TPB) in the literature \citep{zhao_dependence_2005}. The triple phase boundary is the site of intersection of the electrode, the electrolyte and the gas phase. In this work, we use the mass action kinetics detailed chemistry model proposed by \cite{decaluwe_importance_2008}. In this model, the adsorption is modelled though the gas--electrode surface reactions and the gas--electrolyte surface reactions. There is only one edge phase, which represents the TPB. The edge reactions involve only the adsorbed surface phase species. The rate equation (\ref{eq:totalSurfaceReactionRate}) simplifies to,
 \begin{align} \label{eq:totalSurfaceReactionRateTPB_electrode}
      \dot X_{a,\rm electrode}^{\rm (s)} = \frac{1}{\Gamma_{\rm electrode}} \left( \dot M_{a,\rm \rm electrode}^{\rm (s)} +  \frac{1}{a^{(s)}_{\rm electrode}} l^{(e)}_{\rm TPB} \dot M_{a,\rm TPB}^{\rm (e)} \right)
 \end{align}
for the species adsorbed on the electrode material surface and,
 \begin{align} \label{eq:totalSurfaceReactionRateTPB_electrolyte}
      \dot X_{a,\rm electrolyte}^{\rm (s)} = \frac{1}{\Gamma_{\rm electrolyte}} \left( \dot M_{a,\rm \rm electrolyte}^{\rm (s)} +  \frac{1}{a^{(s)}_{\rm electrolyte}} l^{(e)}_{\rm TPB} \dot M_{a,\rm TPB}^{\rm (e)} \right)
 \end{align}
for the species adsorbed on the electrolyte material surface.

The oxidation reactions in the anode and the reduction reactions in the cathode are defined to occur in the TPB. Consequently, the electron production rates are a function of the composition of the adsorbed species on their respective surfaces as well as the potential $\Phi$ in the bulk electrolyte and the bulk electrode phase. The electric current is obtained as a product of the Faraday's constant $\it F$ and the molar production rate of the electron $\dot M_{\rm electron,\rm TPB}^{\rm (s)}$. The volumetric current density $\mathcal{I}^{\rm (v)}$, which is the current generated per unit volume of the REV is calculated as, 
\begin{align} \label{eq:volumetricCurrent}
    \mathcal{I}^{\rm (v)} = \; \it F \; l^{(e)}_{\rm TPB} \dot M_{\rm electron,\rm TPB}^{\rm (e)}. 
\end{align}
A positive $\mathcal{I}^{\rm (v)}$ indicates generation of electrons, which is a result of oxidation while a negative $\mathcal{I}^{\rm (v)}$ is a consequence of destruction of electrons due to a reduction reaction. 

\subsection{Charge transport}
\label{sec:chargeTransport}

A simplest Solid Oxide Fuel Cell (SOFC) membrane electrode assembly (MEA) consists of three major sections, as skeched in Fig. (\ref{fig:sofcSketch}). A porous composite anode section made of nickel and YSZ, an impervious solid electrolyte section consisting only of YSZ and a porous composite cathode section made up of LSM and YSZ \citep{bove_modeling_2006}. In order to model charge transport in the cell, we follow the model proposed by \cite{bessler_new_2007}. The equations therein are reproduced in this section for completeness. A typical MEA consists of very thin layers of the three sections sandwiched together. 
We model the charge transport by only considering the gradients in a direction $x$, which is normal to the interface between these layers. We also colloquially refer to this direction to be along the length of the MEA. 

Since the electrode phase has a very high electron conductivity compared to the electrolyte phase, the electrode phase is assumed to have a spatially constant electric potential, i.e.,
\begin{align} \label{eq:electrodePotential}
    \partial_x \Phi_{\rm electrode} = 0.
\end{align}
The electrolyte phase on the other hand has a finite ion conductivity. The current density per unit area $\mathcal{I}^{\rm (a)}$ is obtained by integrating the volumetric current density over and along length of the MEA. 
\begin{align} \label{eq:areaCurrentDensity}
    \mathcal{I}^{\rm (a)}(x) = \int \mathcal{I}^{\rm (v)}(x) dx.
\end{align}
The $\mathcal{I}^{\rm (a)}$ is often concisely referred to as the current density in the literature. By Ohm's law, the potential in the electrolyte phase $\Phi_{\rm electrolyte}$ is obtained by integrating the product of resistivity of the electrolyte phase $\Omega_{\rm electrolyte}$ and the current density.
\begin{align} \label{eq:electrolytePotential}
    \Phi_{\rm electrolyte} (x) = \int \mathcal{I}^{\rm (a)}(x) \; \Omega_{\rm electrolyte}(x)  dx.
\end{align}
The resistivity of the electrolyte phase is function of space, owing to the different composition of the MEA components. 
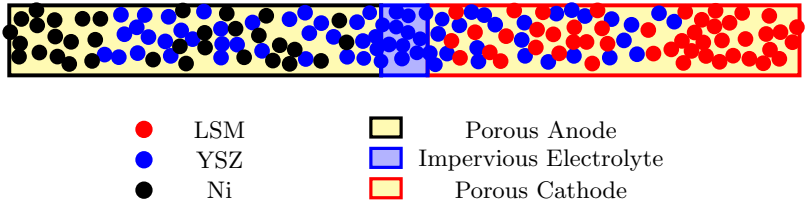
\begin{figure}
	\centering		\includegraphics[width=0.85\linewidth]{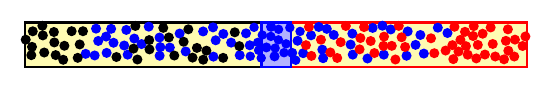} \\
  \begin{tikzpicture}
  \draw [black, thick,fill=black] (0,0) circle [radius=0.1];
  \node at (1,0) {Ni} ;
  \draw [blue, thick,fill=blue] (0,0.4) circle [radius=0.1];
  \node at (1,0.4) {YSZ} ;
  \draw [red, thick,fill=red] (0,0.8) circle [radius=0.1];
  \node at (1,0.8) {LSM} ;
    \draw[red, very thick, fill=yellow, fill opacity=0.3] (3,-0.1) rectangle (3.4,0.15);
  \node at (1+0.75+3+0.5,0) {Porous Cathode} ;
  \draw[blue, very thick, fill=blue, fill opacity=0.3] (3,-0.1+0.4) rectangle (3.4,0.15+0.4);
  \node at (1+0.75+3+0.5,0.4) {Impervious Electrolyte} ;
  \draw[black, very thick, fill=yellow, fill opacity=0.3] (3,-0.1+0.8) rectangle (3.4,0.15+0.8);
  \node at (1+0.75+3+0.5,0.8) {Porous Anode} ;
  \end{tikzpicture}
		\caption{Sketch of a 1 D SOFC.}
		\label{fig:sofcSketch}
\end{figure}
In the simulations, we set a certain potential difference on the anode side of the cell $\Delta \Phi_{\rm anode} = \Phi_{\rm Ni} - \Phi_{\rm YSZ}$. The $\Phi_{\rm YSZ}$ is chosen to be zero at the beginning of the electrochemically active section of the anode. This potential difference along with the local composition of adsorbed species and the temperature, produces a certain volumetric current density described by equation (\ref{eq:volumetricCurrent}). The volumetric current density in the sandwiched solid electrolyte section is zero. Next, we integrate the volumetric current density with equation (\ref{eq:areaCurrentDensity}) to obtain the current density $\mathcal{I}^{\rm (a)}(x)$ and simultaneously integrate the current density with equation (\ref{eq:electrolytePotential}) to obtain the electrolyte potential throughout the length of the active electrolyte phase. 
Once the current density and the electrolyte potential upto the electrolyte section--cathode section interface becomes known by integration, the cathode electrode potential $\Phi_{\rm LSM}$ is determined by solving (\ref{eq:volumetricCurrent}) in a reverse manner. In other words, the potential difference $\Delta \Phi_{\rm cathode} = \Phi_{\rm LSM} - \Phi_{\rm YSZ}$ is varied iteratively by guessing $\Phi_{\rm LSM}$ with the Secant method such that the target current density, matching the same current density as the anode, is obtained. 
The cell voltage is then $\Phi_{\rm cell} = \Phi_{\rm LSM} - \Phi_{\rm Ni} $. 
\subsection{Source terms in the hydrodynamics equations}
\label{sec:sourceTerms}
As defined by equation (\ref{eq:totalFluidReactionRate}), when the mass production rate $\dot \rho_a^{\rm (f)}$ includes the production rates due to adsorption and desorption reactions, the total post collision fluid density of the mixture $\rho^{(pc)}$ changes from the density of the mixture before the species collision step (\ref{eqn:finalNumericalEquationsReactive}) is executed. From the nature of the reaction term (\ref{eq:R2}), the change in fluid density can be computed by summing over the mass production rates,
\begin{align} \label{eq:rho_pc}
     \phi \rho^{(pc)} &=  \phi \sum_a^M \rho_a + \sum_a^M \dot \rho_a^{\rm (f)} \delta t &&=  \phi \sum_a^M \rho_a + \sum_a^M \left( \dot \rho_a^{\rm r} +   \sum_k^\kappa a^{(s)}_k \dot \rho_{a,k} \right) \delta t. \\
     &
     &&=  \phi \rho  
    + \sum_a^M \sum_k^\kappa a^{(s)}_k \dot \rho_{a,k}  \delta t. 
\end{align}
The new density $\rho^{(pc)}$ is recorded by the species system of equations (\ref{eqn:finalNumericalEquationsReactive}) due to the presence of the source terms. However, due to the reduced description of the model, the kinetic equations for the momentum (\ref{eqn:f}) need to be informed about this change. Analogous to the change in density, there are also implications for the energy tracked by equations (\ref{eqn:g}), which we deal with in this section.

In this work, we assume that the adsorbed species and the fluid phase species are in thermal equilibrium. Therefore, an ``unified" internal energy $\tilde U$ can be defined in a REV as a result of the sum of internal energies of both the fluid state species and the adsorbed species. If there exists $M$ fluid state species, $B$ adsorbed species on the electrolyte surface and $D$ adsorbed species on the electrode surface, the unified internal energy at a temperature $T$ is given by,  
\begin{align}
\label{eq:temperatureCell}
     \tilde U = &  \phi \rho \sum_{a=1}^MY_a\left[U_a^0 + \int_{T_0}^T {C}_{a,v}(T')dT'\right]  \nonumber \\ 
    & + a^{(s)}_{\rm electrolyte} \rho^{(s)}_{\rm electrolyte} \sum_{b=1}^B Y_b\left[U_b^0 + \int_{T_0}^T {C}_{b,v}(T')dT'\right] \nonumber \\
    & + a^{(s)}_{\rm electrode} \rho^{(s)}_{\rm electrode} \sum_{d=1}^D Y_d\left[U_d^0 + \int_{T_0}^T {C}_{d,v}(T')dT'\right].
\end{align}
Electric power produced due to the production of electrons is lost from the hydrodynamic system \citep{kee_chemically_2017}. This electric power can be calculated based on the local volumetric current as,
\begin{align}
    \tilde P_{\rm elec} = \lvert \Phi_{electrode} - \Phi_{electrolyte} \rvert  \; \mathcal{I}^{\rm (v)}
\end{align}
The resistance of the electrolyte phase to the flow of electrons leads to heating of the hydrodynamic system. This ohmic heat rate is given by,
\begin{align}
    \tilde P_\Omega = \left( \frac{a_{\rm electrolyte}}{ l^{(e)}_{\rm TPB} } \right) \mathcal{I}^{\rm (v)}  \mathcal{I}^{\rm (v)} \Omega \delta x  
\end{align}
The net effect of the energy lost as electrical work and the heat gained due to the electrolyte phase resistance is combined into an energy source term $\tilde U_{\rm source}$,
\begin{align}
    \tilde U_{\rm source} = ( - \tilde P_\Omega + \tilde P_{\rm elec} ) \delta t 
\end{align}
Therefore, the unified internal energy post the species collision step  is,
\begin{align}
     \tilde U^{(\rm pc)} = \tilde U + \tilde U_{\rm source}  
\end{align}
After the collision step, we need to know the post collision internal energy $\rho U^{(\rm pc)}$ of the fluid state, since the kinetic equations only track the evolution of the bulk fluid phase. The  post collision internal energy is found by simple arithmetic manipulation of subtracting the post collision energy of the adsorbed phases from the post collision unified internal energy.  
\begin{align}
\label{eq:InternalEnergyCell_PC}
    \phi \rho^{(\rm pc)} U^{(pc)} = \tilde U^{(\rm pc)} &
    - a^{(s)}_{\rm electrolyte} \rho^{(s)(\rm pc)}_{\rm electrolyte} \sum_{b=1}^B Y_b^{(\rm pc)} \left[U_b^0 + \int_{T_0}^{T} {C}_{b,v}(T')dT'\right] \nonumber \\
    & - a^{(s)}_{\rm electrode} \rho^{(s)(\rm pc)}_{\rm electrode} \sum_{d=1}^D Y_d^{(\rm pc)} \left[U_d^0 + \int_{T_0}^{T} {C}_{d,v}(T')dT'\right].
\end{align}
%
%
%
%
%
The post species collision internal energy $U^{(\rm pc)}$ is used to generate an energy distribution $g_i^{\rm eq, source}$ analogous to (\ref{eqn:g0momTotalEnergy}), with its zeroth moment containing the new internal energy, 
\begin{align} \label{eqn:g0mom_source}
& \sum_{i=0}^{Q-1} g_i^{\rm eq, source} =  \phi \rho^{\rm (pc)} \left( U^{\rm (pc)} + {\frac{u^2}{2 {\phi^2}}} \right),   
\end{align}
Finally, the kinetic equations (\ref{eqn:f}) and (\ref{eqn:g}) are appended to account for the mass and energy changes,
\begin{align} 
f_i(\bm{x}+\bm{c}_i \delta t,t+\delta t)- f_i(\bm{x},t)&=  \omega (f_i^{\rm ex} -f_i) + \left( \frac{\rho^{(pc)}}{\rho} f_i^{\rm ex} - f_i^{\rm ex} \right), \label{eqn:f_source} 
\\
g_i(\bm{x}+\bm{c}_i \delta t,t+ \delta t) - g_i(\bm{x},t)&=  \omega_1 (g_i^{\rm eq} -g_i) + (\omega - \omega_1) (g_i^{*} -g_i) + (g_i^{\rm eq,source}-g_i^{\rm eq}).
 \label{eqn:g_source}
\end{align}
The kinetic equations for species (\ref{eqn:finalNumericalEquationsReactive}) already include populations $\dot f_{ai}^{\rm r}$ for mass sources through (\ref{eq:R2}). In case of both homogeneous and heterogeneous reactions, those populations are computed with the total mass production rate $\dot \rho_a^{\rm (f)}$,   
\begin{align}
    	\dot f_{ai}^{\rm r}
    	&= \dot \rho_a^{\rm (f)}  \Psi_{c_{ix}}\left(\frac{u_{x}}{{\phi}},\frac{u_{x}^2}{{{\phi^2}}}+R_aT\right)
	\Psi_{c_{iy}}\left(\frac{u_{y}}{{\phi}},\frac{u_{y}^2}{{\phi^2}}+R_aT\right) 
	\Psi_{c_{iz}}\left(\frac{u_{z}}{{\phi}},\frac{u_{z}^2}{{\phi^2}}+R_aT\right).
\label{eq:fsourceHetero} 
\end{align}
}
\section{Validation}
Together in sections (\ref{sec:stefanMaxwell}) and (\ref{sec:lbMixtures}), a lattice Boltzmann model has been formulated for computing multicomponent hydrodynamics in porous media. To numerically verify the resultant formulation which been theoretically shown to replicate the dusty gas model at the macroscopic scale, we evaluate the component diffusion fluxes in a ternary mixture undergoing counter diffusion through a capillary tube. In further development which has been described in section (\ref{sec:electrodeModel}), the model has been made capable of heterogeneous reactions and electrochemistry.  The further development is validated by comparing the polarization curves of a SOFC against LBM simulation of a membrane electrode assembly consisting of porous composite electrodes.  

\subsection{Counter diffusion in capillary tubes}
\label{sec:dgm}
\begin{figure}
    \centering
    \includegraphics[width=0.5\linewidth]{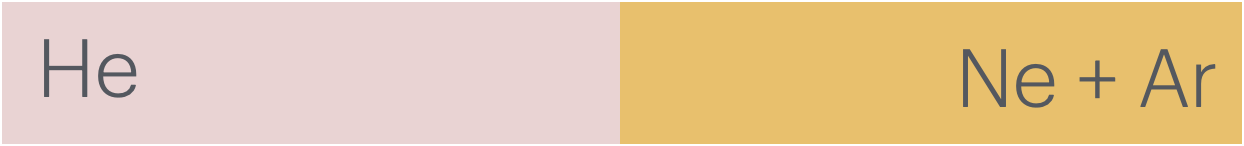}
    \caption{Sketch of the initial conditions in the capillary tube.}
    \label{fig:capillaryTube}
\end{figure}
\begin{figure}
    \centering
    \includegraphics[width=1.0\linewidth]{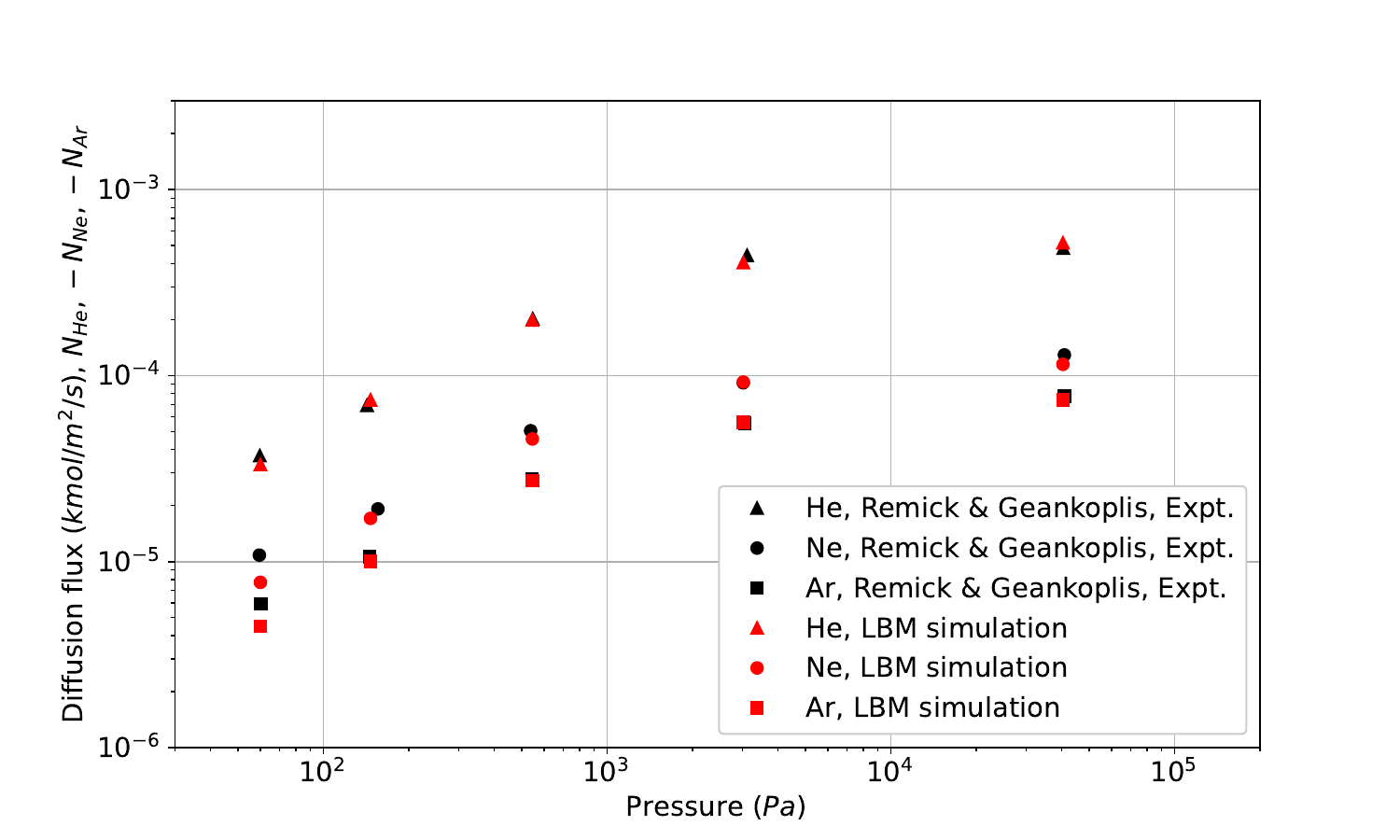}
    \caption{Molar flux of helium counter diffusing against neon and argon  in a $39 \mu m$ wide and $9.6 mm$ long capillary tube. One dimensional LBM simulations compared with experiments from \cite{remick_ternary_1974} performed at different pressures ranging from $60 Pa$ to $40422 Pa$. }
    \label{fig:dgmFlux}
\end{figure}
In order to validate the dusty gas model, a one dimensional domain of $1920$ lattice nodes is used to represent a $9.6mm$ capillary tube. Following the experiments of \cite{remick_ternary_1974}, the validation set consists of $5$ simulations, each differing in the initialised spatially uniform pressure but having the same spatially uniform initial temperature of $300 \rm K$. Across the $5$ individual tests, the pressure is varied over a wide range, from the almost completely molecular diffusion regime at $40422 \rm Pa$ to the nearly fully Knudsen diffusion dominated regime at $60 \rm Pa$. For all $5$ simulations, the left half of the simulation domain is approximately initialized with $\{X_{\ce{He}}=0.9585,X_{\ce{Ne}}=0.0265,X_{\ce{Ar}}=0.0150\}$, while the right half of the domain is approximately initialised with $\{X_{\ce{He}}=0.0571,X_{\ce{Ne}}=0.5125,X_{\ce{Ar}}=0.4304\}$. To reproduce the experiments faithfully, small variations in the mole fractions appearing at the ends of the capillary tubes in the experiment are accounted for by following the exact values presented in Table 1 of \cite{remick_ternary_1974}. 

The ends of the computational domain are supplied from inlets with a mixture having the same composition as the one they were initialized with. The temperature of the incoming mixture is $300 \rm K$ and the pressure is equal to the initial pressure corresponding to the test case. The inlet boundary condition is imposed by replacing only the missing populations, which has been described in detail as a `simplified flux boundary condition' in \cite{sawant_consistent_2022}. In order to get the binary diffusion coefficients $\mathcal{D}_{ab}$ from the Cantera \citep{goodwin_cantera_2018} package, it has been supplied with the parameters pertaining to Lennard-Jones potential for helium, neon and argon based on \cite{tchouar_thermodynamic_2003}, \cite{klopper_highly_2001,nasrabad_prediction_2004} and \cite{tee_molecular_1966}, respectively. For a capillary tube with diameter $d_0=39 \mu m$ the Knudsen diffusion coefficients $\mathcal{D}_{a}^{\rm k}$ \citep{mason_gas_1983} are calculated as,
\begin{align}
    \mathcal{D}_{a}^{\rm k} = \frac{\phi}{\bar \tau} \frac{d_0}{3} \sqrt{\frac{8}{\pi} \frac{R_u T}{m_a}}.
\end{align}
For this specific test case, since the capillary tube is fully open, the porosity $\phi$ is one and the tortuosity $\bar \tau$ is also unity because the capillary tube is straight. Since the tube diameter, the porosity and the tortuosity are the only parameters from the physical setup which are entered into the model, it is an excellent candidate to test a formulation for dusty gas diffusion model because this test case does not involve any tunable free parameters. For the purpose of comparison, the component mass diffusion fluxes $\rho_a \bm{V_a}$ obtained through equations (\ref{eq:density1}) and (\ref{eqn:transform1Alternate}) are converted to physical units and divided by their molecular weights $m_a$ to get the molar diffusion fluxes $N_a$. At the steady state, the molar diffusion flux of helium, and the negative of the molar diffusion flux of neon and argon which flows in the opposite direction is plotted in Fig. (\ref{fig:dgmFlux}). The fluxes from the lattice Boltzmann simulations compare well with the fluxes reported from experiments in \cite{remick_ternary_1974}. The model has reproduced the correct flux in the molecular diffusion regime at the higher pressures, the transition regime at moderate pressures,  as well as at the Knudsen diffusion dominated regime at low pressures. This test instills enough confidence that the dusty gas model has been correctly formulated and realised in the lattice Boltzmann framework.
%
%
\subsection{Solid Oxide Fuel Cell Simulation}
\label{sec:sofc}
\begin{table}
    \centering
    \begin{tabular}{| c | c | c}
        Phase  & Species  & \\ \hline
        Gas bulk & \ce{O2}, \ce{H2}, \ce{H2O}, \ce{N2}, \ce{Ar} & \\
        Anode side electrode surface & $\ce{Ni}_{(\ce{Ni})}$, $\ce{H}_{(\ce{Ni})}$, $\ce{H2O}_{(\ce{Ni})}$, $\ce{O}_{(\ce{Ni})}$, $\ce{OH}_{(\ce{Ni})}$ & \\
        Anode side electrolyte surface & $\ce{O}_{(\ce{YSZ})}$, $\ce{OH}_{(\ce{YSZ})}$, $\ce{H2O}_{(\ce{YSZ})}$, $\ce{YSZ}_{(\ce{YSZ})}$ & \\
        Cathode side electrolyte surface & $\ce{O}_{(\ce{YSZ})}$, $\ce{YSZ}_{(\ce{YSZ})}$ & \\
        Cathode side electrode surface & $\ce{O}_{(\ce{LSM})}$, $\ce{LSM}_{(\ce{LSM})}$ & \\
        Electrode bulk & $\ce{e^-}_{\ce{Ni}(\rm b)}$ & \\
        Electrolyte bulk & $\ce{O^{2-}}_{\ce{YSZ}(\rm b)}$ & \\
    \end{tabular}
    \caption{List of species in respective phases as defined in the chemical mechanisms from \cite{decaluwe_importance_2008}.}
    \label{tab:listOfSpecies}
\end{table}
\begin{table}
    \centering
    \begin{tabular}{| c | c | c | c |}
        Porosity $\phi$ & Utilization thickness $\delta_{\rm util} (\mu m)$  & $l_{\rm TPB}^{(e)} (m/m^3)$  & Resolution $\delta x$ $(\mu m)$ \\ \hline
        $0.57$ & $2.0$ & $3.00 \times 10^{13} $ & $1.0$    \\
        $0.48$ & $3.0$ & $0.80 \times 10^{13} $ & $1.5$ \\
        $0.32$ & $5.0$ & $0.14 \times 10^{13} $ & $2.5$ \\
    \end{tabular}
    \caption{Parameters corresponding to simulations performed at different porosities.}
    \label{tab:listOfParameters}
\end{table}
\begin{figure}
    \centering
    \includegraphics[width=0.98\linewidth]{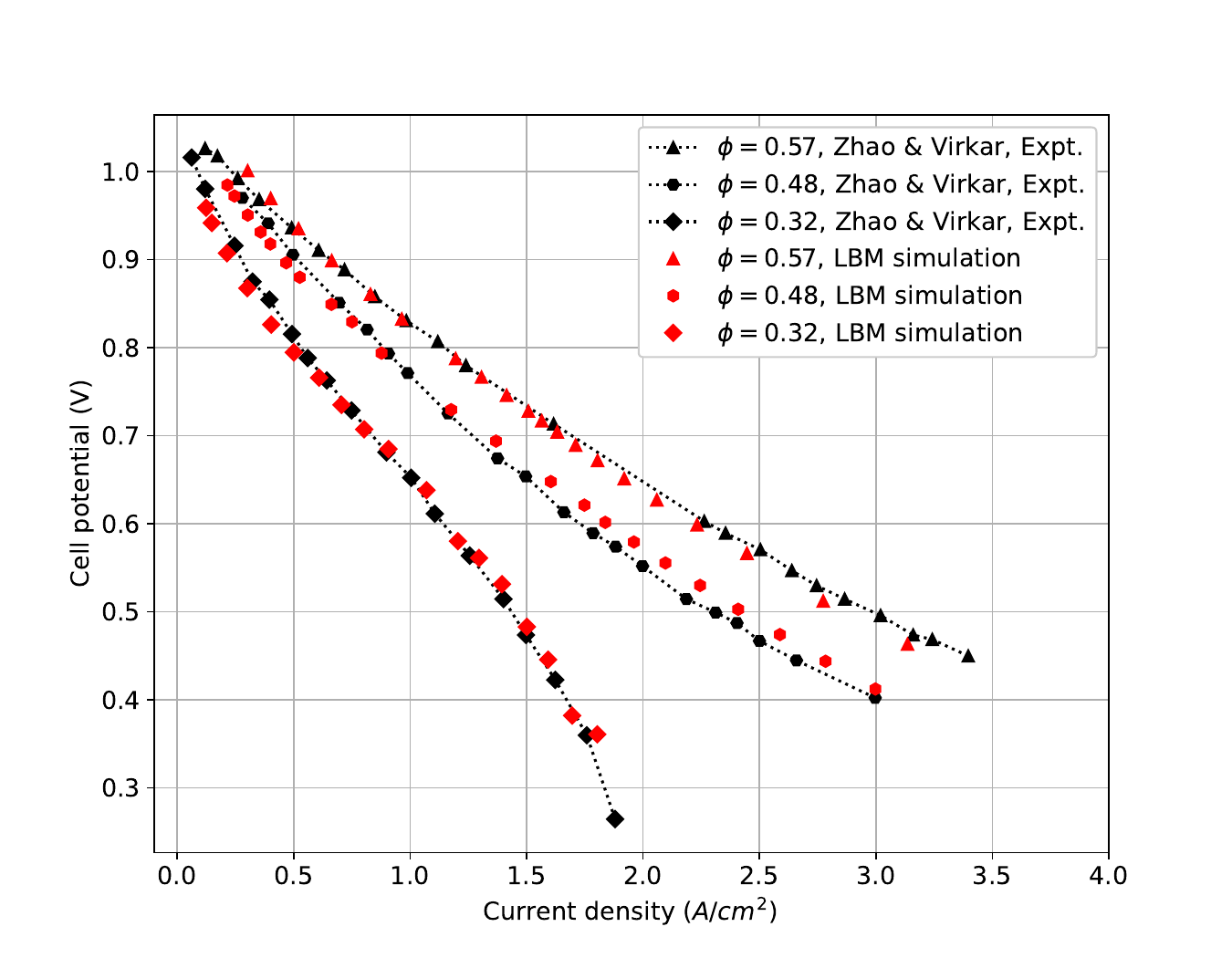}
    \caption{Cell potential vs.\ current density.}
    \label{fig:voltageCurve}
\end{figure}
\begin{figure}
    \centering
    \includegraphics[width=0.98\linewidth]{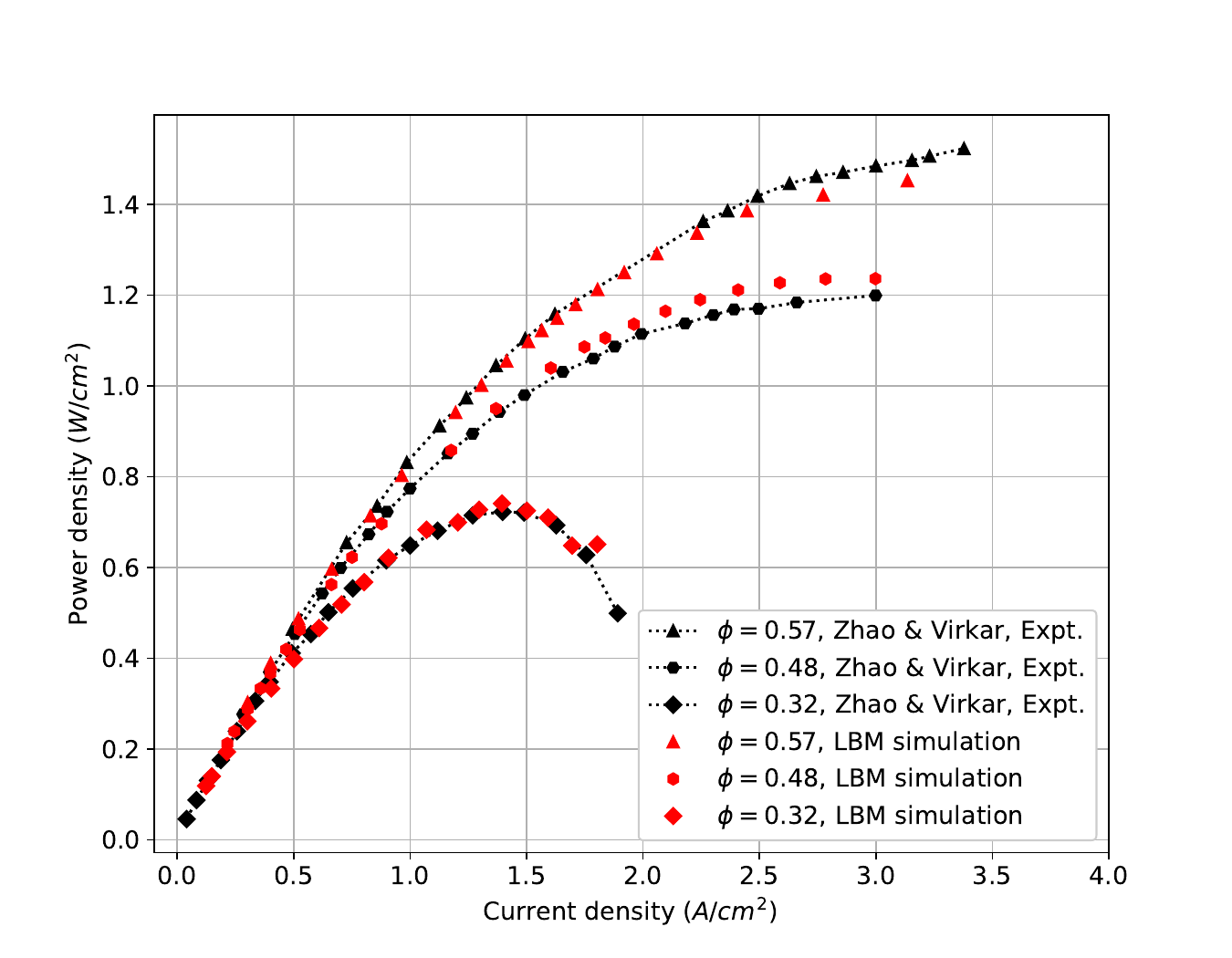}
    \caption{Power density vs.\ current density.}
    \label{fig:powerCurve}
\end{figure}
\begin{figure}
    \centering
    \includegraphics[width=0.85\linewidth]{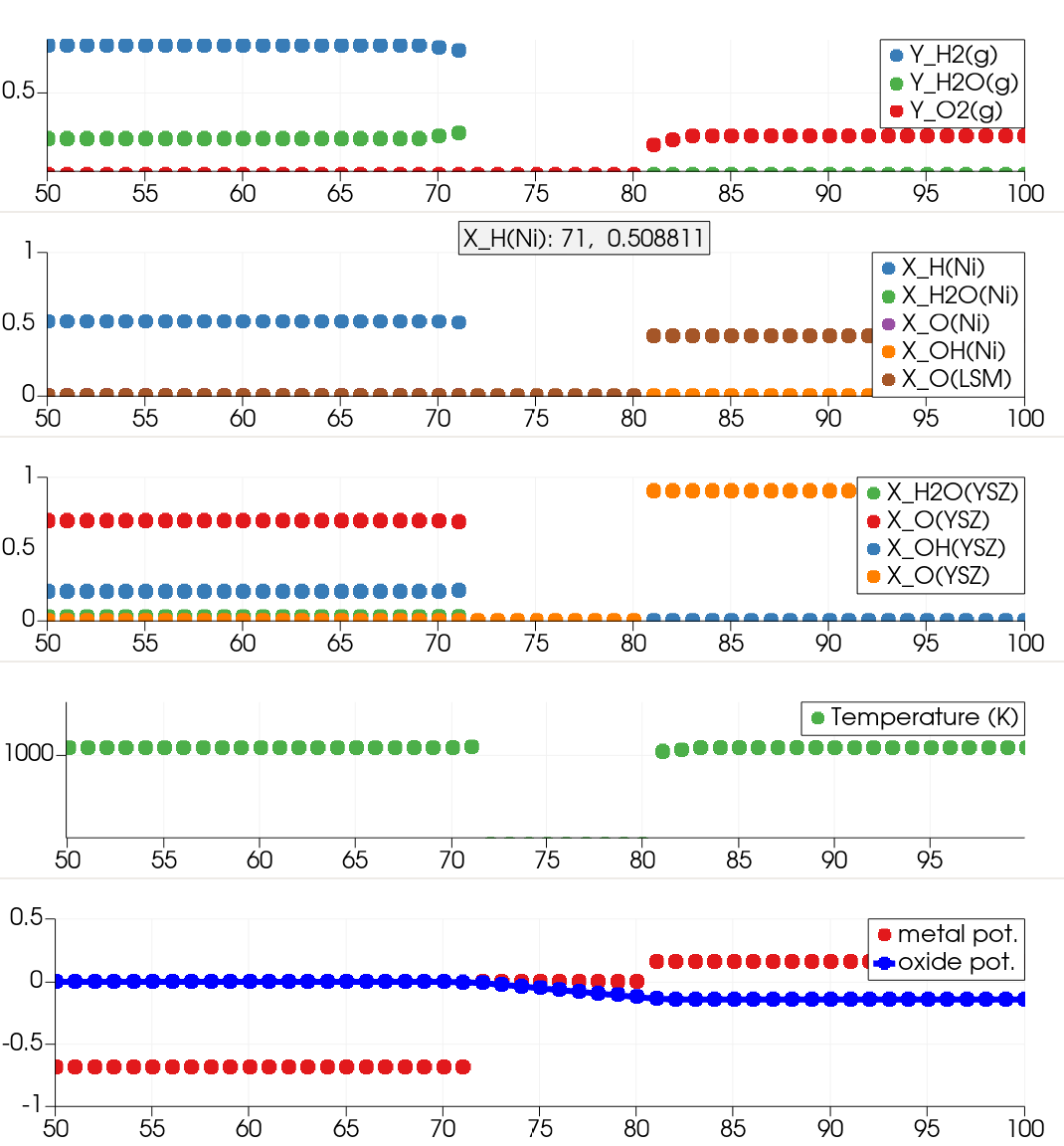}
    \caption{Top to bottom: Mass fraction of the gas phase species, mole fraction of the species adsorbed on the electrode surface, mole fractions of the species adsorbed on the electrolyte surface, temperature and potential along the length of the MEA. Porosity is $\phi=0.57$ and the current density is $1.05 \; A/cm^2$. }
    \label{fig:profilesSpecies}
\end{figure}
\begin{figure}
    \centering
    \includegraphics[width=0.85\linewidth]{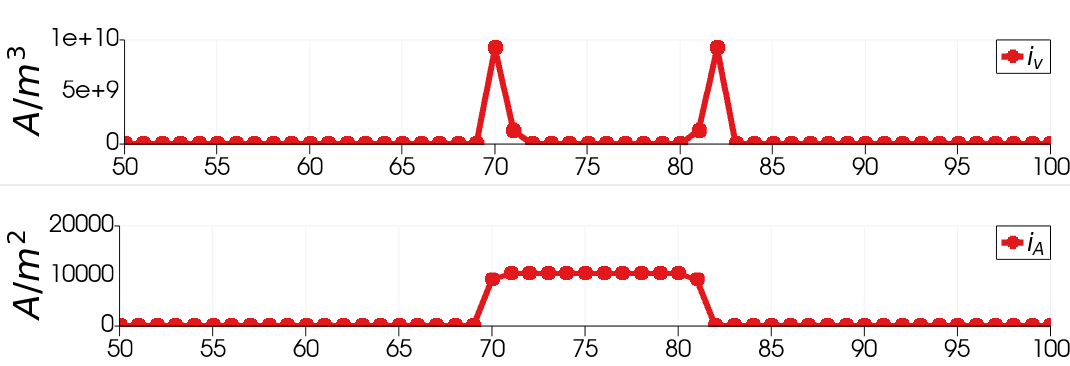}
    \caption{Top to bottom: Current density per unit volume $\mathcal{I}^{\rm (v)}(x)$ and current density per unit area $\mathcal{I}^{\rm (a)}(x)$ along the length of the MEA. Porosity is $\phi=0.57$ and the current density is $1.05 \; A/cm^2$.}
    \label{fig:profilesCurrents}
\end{figure}
\begin{figure}
    \centering
    \includegraphics[width=0.85\linewidth]{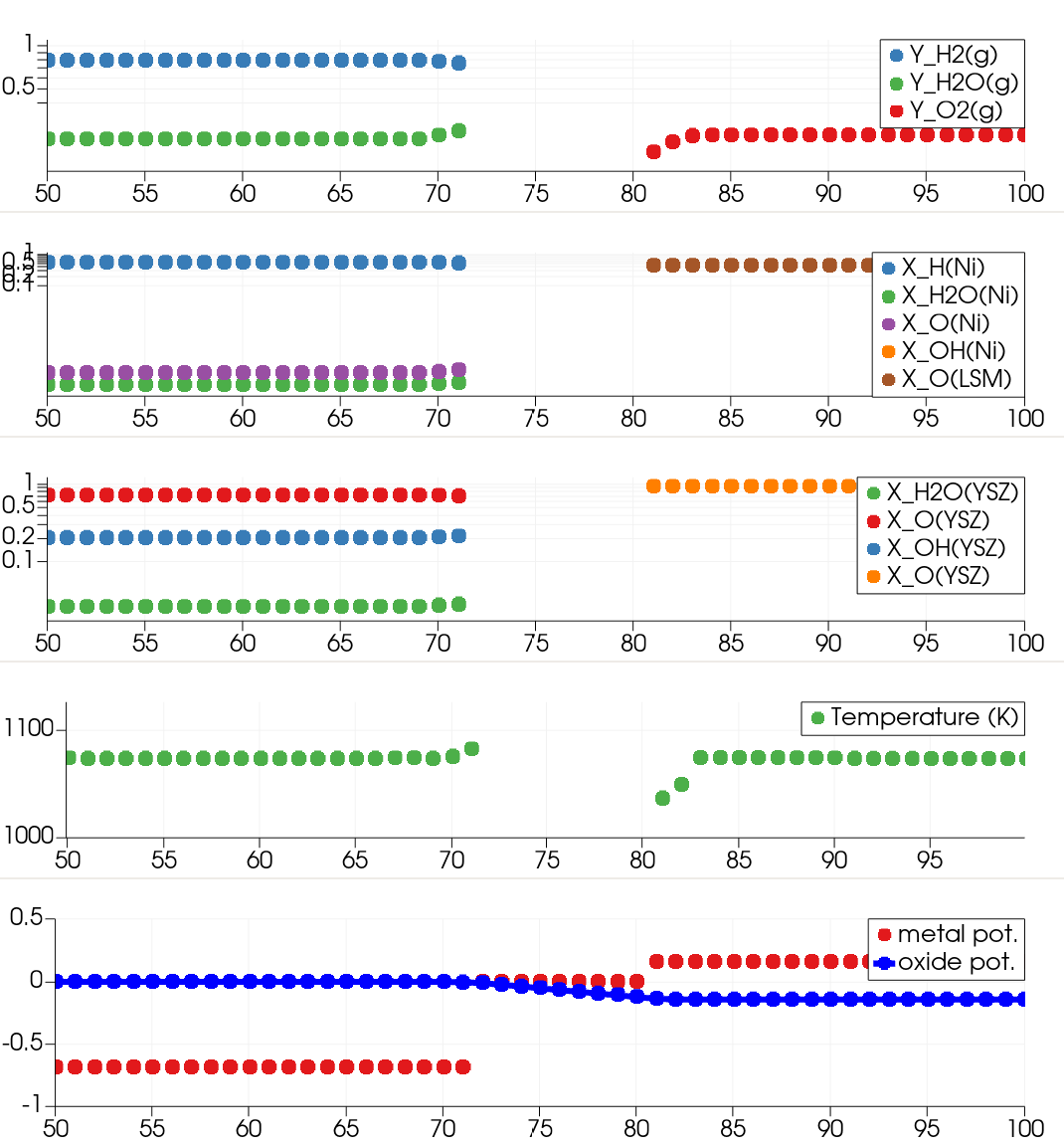}
    \caption{Top to bottom in log scale: Mass fraction of the gas phase species, mole fraction of the species adsorbed on the electrode surface and mole fractions of the species adsorbed on the electrolyte surface. In linear scale: Temperature and potential along the length of the MEA. Porosity is $\phi=0.57$ and the current density is $1.05 \; A/cm^2$. }
    \label{fig:profilesSpecies_log}
\end{figure}
The LBM is used to perform $\rm 1D$ simulations of a SOFC member electrode assembly at different porosities and current densities with an intent to obtain polarization curves verifiable against the literature. As described in section (\ref{sec:StandardLattice}), we use the standard $D3Q27$ lattice albeit with periodic boundary conditions in two directions in order to achieve a one dimensional computational domain. We recall that since we are working within the REV formulation, a one dimensional computational domain sufficiently accounts for the three dimensional microstructure of the porous electrodes through parameters such as the porosity, specific areas and specific lengths. A sketch of the simulation domain is shown in Fig (\ref{fig:sofcSketch}) to aid visualisation of the geometry. The length of the impervious electrolyte section in the middle is $8 \;\mu m$, while the total length of the cell is $150 \;\mu m$. 

We use the electrochemical and heterogeneous reaction mechanisms from \cite{decaluwe_importance_2008}, consisting of $5$ gas phase species, namely \ce{O2}, \ce{H2}, \ce{H2O}, \ce{N2} and \ce{Ar}. Besides the gas phase species, the mechanism also contains species which exist as adsorbed species on the surfaces of the anode material $_{(\ce{Ni})}$, cathode material $_{(\ce{LSM})}$ and the electrolyte material $_{(\ce{YSZ})}$. All the species are listed in Table (\ref{tab:listOfSpecies}). Their mechanism is a compilation of anode reactions from \cite{bessler_influence_2007}, cathode reactions from \cite{jiang_electrochemical_1998}, charge transfer reactions from \cite{singhal_solid_2005} and thermodynamic parameters from various other sources \citep{gordon-1994,mcbride-1996,bessler_new_2007,janardhanan_cfd_2006}.

Apart from the gas phase and the adsorbed surface phase species, the electrons $\ce{e^-}_{\ce{Ni}(\rm b)}$ reside in the bulk electrode phase while the oxide ions $\ce{O^{2-}}_{\ce{YSZ}(\rm b)}$ reside in the bulk of the electrolyte phase. In this paper, we do not model the transport inside these solid bulk phases. In the unit of mole fractions, the anode is uniformly initialised with a composition of \{\ce{H2} 0.97, \ce{H2O} 0.03\}, whereas the cathode is uniformly initialised with a composition of \{\ce{O2} 0.21, \ce{N2} 0.78, \ce{Ar} 0.01\}. Throughout the simulation, the electrodes are supplied from the inlets with a mixture having the same composition as the one they were initialized with. An inlet boundary condition is used at both ends of the computational domain to supply the respective mixtures at a pressure of one atmosphere and a temperature of $1073.15 \rm K$. The inlet boundary condition is imposed by only replacing the missing populations, which has been described in detail as a `simplified flux boundary condition' in \cite{sawant_consistent_2022}. At the interface of the porous electrode section with the impervious electrolyte section, bounce back boundary condition is applied on the missing populations of the species lattice as well the on the missing populations of the mean field momentum lattice and the energy lattice. At the macroscopic level, the application of the bounce back boundary condition results into no flux, no slip and adiabatic conditions at the interface \citep{he_novel_1998}. 

For the purpose of validation, we aim to replicate polarization curves from the experiments performed on  solid oxide button cells in \cite{zhao_dependence_2005}. In \cite{decaluwe_importance_2008}, the curves have been reproducible through a finite volume DGM discretization in conjunction with a set of detailed reaction mechanisms, the latter of which is also used in this work in conjunction with the proposed LBM. To that end, simulations are performed for three values of anode porosity, $0.57$, $0.48$ and $0.32$, while the cathode porosity is kept constant at $0.45$. In reality, the length of the electrochemically active layer in the porous electrodes is determined by the transport of the $\ce{O^{2-}}_{\ce{YSZ}(\rm b)}$ ions through the bulk of the electrolyte. Since we do not model the ion transport through the bulk electrolyte, it is not possible to know the length of the electrochemically active layer in the porous electrodes. Consistent with other studies performed under the same limitations, including that of \cite{decaluwe_importance_2008}, the length of the active layer, also called the `utilization thickness' $\delta_{\rm util}$ as well as the length of the triple phase boundary $l_{\rm TPB}^{(e)}$ are treated as a free parameter. Their values used in the LBM simulations are reported in Table (\ref{tab:listOfParameters}). The utilization thickness has an inverse relation to porosity as expected \citep{divisek_structure_1999,chan_anode_2001}. The size of the utilization thickness in the simulation is controlled by changing the resolution. 
A lattice resolution of $\delta x = 1\mu m$ is used for $\phi=0.57$ simulation and the resolution is changed to vary the utilization length. The product of utilization thickness and specific surface area $\delta_{\rm util} a^{(s)}$ is $20$ for all surfaces and all simulations. 

Although we do not model the transport of individual ions, we solve the charge transport equations (\ref{eq:areaCurrentDensity}) and (\ref{eq:electrolytePotential}) in the solid electrolyte including in the middle impervious section, using the values of resistivity from \cite{zhao_dependence_2005}. Therefore, the middle impervious electrolyte section does cause a drop in the cell potential due to resistance of the electrolyte phase. The electrolyte phase in the anode section is set to $0$ Volts and the metal electrode phase is set at a certain potential above the open circuit voltage so that a certain current density is produced in the anode. This current density is then imposed at the cathode and a reverse problem is solved, i.e. an electrode potential which produces the same current is found iteratively by the Newton--Raphson method. 

The polarization curves obtained from the simulations have been compared against the experimental data produced by \cite{zhao_dependence_2005}. Figure (\ref{fig:voltageCurve}) compares the voltage vs. current density whereas Fig. (\ref{fig:powerCurve}) compares power vs. the current density. From both the figures, it is evident that a reasonable match has been obtained with respect to the experimental data for all three values of porosities in the regime of low as well as medium and high current densities. The model captures losses in all three regimes: the activation overpotential occuring at low current density originating from activation of the electrochemical reactions, the ohmic overpotential occuring at moderate current density originating from the loss due to internal resistance and the concentration overpotential occuring at high current density due to limitation on the transport of the reactants in the porous electrodes. The effect of concentration overpotential is unmistakably visible in the polarization curve for $\phi=0.32$ after the current density $1.5 \; A/cm^2$. Good agreement with the experiments is an indication that modeling of the chemistry, charge continuity and most importantly the species transport with associated momentum and energy modeling has been achieved correctly. 
In our future work, we intend to implement not only tortuosity as a parameter but also model the ion transport in the bulk electrolyte which will make it possible to obtain the utilization length as an output from the simulation rather than as an input parameter. 

Since simulations with detailed chemistry as well as hydrodynamics have been performed, there is an opportunity to peek into the porous electrodes in order to observe the state of various variables that have been modeled. In Fig. (\ref{fig:profilesSpecies}), a data point has been selected for plotting the composition, temperature and potential along the length of the composite electrodes. The data point corresponds to a $\phi=0.57$ cell while it produces approximately $1 A/cm^2$. In the figure, the gas phase composition is seen to have an appreciable change only near the interfaces with the impervious electrolyte. On the left side of the figure, which is the anode section, gaseous $\ce{H2}$ is seen to have a drop in mass fraction accompanied by a rise in the gaseous $\ce{H2O}$ in the topmost frame. This is an indication of the oxidation of hydrogen by the oxide ions, resulting into water. On the cathode section on the right hand side, there is a drop in to mass fraction of $\ce{O2}$, as it gets converted into oxide ions. The figure also shows mole fractions of adsorbed species and temperature, which do not show appreciable change in the linear scale. In the bottom most frame of the figure, the drop in potential due to the resistance of the electrolyte phase is visible across the impervious electrolyte section, labelled as `oxide pot.'. The metal phase potential is negative on the anode side of the cell and positive on the cathode side of the cell as expected. In the accompanying figure (\ref{fig:profilesCurrents}), the top frame shows a peak in the volumetric current density due to the generation of electrons at the anode-electrolyte section interface and a second peak due the consumption of electrons at the cathode-electrolyte section interface. We plot the current as positive even though the rate of production of electrons is negative in the cathode. The profile in the lower frame of Fig. (\ref{fig:profilesCurrents}) is that of the area current density, which is an integrated quantity. In the simulation, the current density generated in anode is imposed as a boundary condition on the cathode and a corresponding cathode voltage is computed. The dropping off of the current density in the cathode section shows that the cell is correctly charge balanced and `connected', i.e. all the electrons created by the electrochemical reactions in the anode are consumed by the electrochemical reactions in the cathode. 

Figure (\ref{fig:profilesSpecies_log}) presents the same information as in Fig. (\ref{fig:profilesSpecies}) with a change in the $Y-$ axis scaling. The mass and the mole fractions are plotted in the log scale to make the species with low mole fractions more visible. For example, in the second sub figure from the top, the adsorbed oxygen and adsorbed water on the nickel surface in the anode section can be seen to have a small increase in the mole fraction at the interface with the electrolyte section. There is a similar increase at the interface in the oxygen and the hydroxide adsorbed on the YSZ surface in the anode, visible in the third sub figure from the top. Perhaps the most interesting part of Fig. \ref{fig:profilesSpecies_log} is the temperature profile. The oxidation reaction in the anode being exothermic is seen to have caused an increase in the temperature whereas the endothermic oxidation \citep{xia_oxygen_2012} in the cathode is seen to have produced a drop in the temperature in the electrochemically active region. This temperature dynamic reveals an interesting opportunity to also compute the heat conduction in the electrolyte phase, which will be undertaken as part of our future work.
\section{Conclusion}
\label{sec:conclusion}
%
%
In this work, we started with a reactive multicomponent compressible lattice Boltzmann model with Stefan-Maxwell diffusion. In section (\ref{sec:stefanMaxwell}), we converted the Stefan-Maxwell diffusion model to the dusty gas model. Next, the hydrodynamic model was rewritten to obtain a representative elementary volume level hydrodynamic model for porous media in section (\ref{sec:lbMixtures}). The resulting dusty gas model was validated by computing counter diffusion molar fluxes through capillary tubes in section (\ref{sec:dgm}). The model was fortified with heterogeneous surface and electrochemical reactions in section (\ref{sec:electrodeModel}). The complete LBM for porous electrochemical flows hence constructed was used to perform simulations of flow in solid oxide fuel cell electrodes and polarization curves were compared against experiments in section (\ref{sec:sofc}).

To recap, through this work, we have developed a lattice Boltzmann model which encompasses detailed electrochemistry, multicomponent mass transport and fluid dynamics for realistic simulations of fuel cells. Such a model is not only a starting point to reveal the rich interplay between chemical reactions, fluid flow, species composition and current density at the pore scale but also has a potential to function as a digital twin for developing and optimizing practical fuel cell microstructures and geometries at a reasonable compute cost. In a reduced form, the model can also be used to study single component or non reactive flows in porous media, which is of much relevance to the geophysics community \citep{philip_flow_1970,winter_mean_2000}. Since the model at it's core is an extension of a compressible lattice Boltzmann model to porous media, the model also has other potential applications including but not limited to petroleum engineering \citep{zidane_efficient_2014,jonsson_fluid_1992}. 

Going forward, we intend to reduce the free parameters in the model by accommodating more physics. To that end, a model for ion transport through the solid electrolyte will set the solver free from having to use the utilization thickness as a parameter. Heat conduction through the solid matrix has to be modeled, which can provide additional opportunities to study the thermal gradients and hence the stresses in the membrane electrode assembly. In addition to the conjugate heat transfer, we also need to account for the capacitance when dealing with the model for electric current if we need to make the model useful for transient simulations. Creation of such a predictive tool is expected to yield theoretical, scientific and practical gains. Mesoscale simulations on full $\rm 3D$ geometries would provide additional insights into flow, composition and temperature fields in typical fuel cells and therefore improve our knowledge of physical and chemical processes occurring in fuel cells. 
A computing framework capable of performing realistic scale simulations of fuel cells will open up interesting opportunities for optimizations pertaining to geometry, operating conditions, fuel composition, internal reforming, etc.

\noindent {\bf Acknowledgement}. 
This work was supported by the Swiss National Science Foundation Proposal `10.001.232 Mesoscale multiphysics modeling for fuel cells' and European Research Council (ERC) Advanced Grant No. 834763-PonD. 
Computational resources at the Swiss National  Super  Computing  Center  CSCS  were  provided  under grant No. s1286. NS thanks Steven DeCaluwe at the Colorado School of Mines for the inputs provided by him on the Cantera Users Group.

\noindent {\bf Declaration of interests}.
The authors report no conflict of interest.
\appendix 
\section{Hydrodynamic limit of the mean-field LBM}
\label{sec:ceNavierStokes}
We expand the lattice Boltzmann equations (\ref{eqn:f}) and (\ref{eqn:g}) in Taylor series to second order, using space component notation and summation convention,
\begin{align}
\left[\delta t (\partial_t + c_{i\mu} \partial_\mu ) + \frac{\delta t^2}{2} (\partial_t + c_{i\mu} \partial_\mu )^2\right] f_i &= { \omega_{\rm B} (f_i^{\rm eq} -f_i^{\rm B}) + \omega (f_i^{\rm B} -f_i) },
\label{eqn:ftaylor} 
\\
\left[\delta t (\partial_t + c_{i\mu} \partial_\mu ) + \frac{\delta t^2}{2} (\partial_t + c_{i\mu} \partial_\mu )^2\right] g_i &= \omega_1 (g_i^{\rm eq} -g_i) + (\omega - \omega_1) (g_i^* -g_i).
\label{eqn:gtaylor}
\end{align}
To obtain the right hand side of (\ref{eqn:ftaylor}) from (\ref{eqn:f}), the substitution $f_i^{\rm ex} = f_i^{\rm B}(1-(\omega_{\rm B}/\omega)) + f_i^{\rm eq} (\omega_{\rm B}/\omega)$ has been used. Such splitting of the extended equilibrium $f_i^{\rm ex}$ into a non-equilibrium $f_i^{\rm B}$ and an equilibrium distribution is only done to ease mathematical analysis of the system. The extra relaxation $\omega_{\rm B}$ introduced here vanishes after converting the equation back to the form of $f_i^{\rm ex}$.  
With a time scale $\bar t$ and a velocity scale $\bar c$, the non-dimensional parameters are introduced  as follows,
\begin{equation}
t'=\frac{t}{\bar t}, \; c_{\alpha}'=\frac{c_{\alpha}}{\bar c}, \; x_{\alpha}'=\frac{x_{\alpha}}{\bar c \bar t}.
\label{eqn:nd} 
\end{equation}
Substituting the relations (\ref{eqn:nd}) into  (\ref{eqn:ftaylor}) and (\ref{eqn:gtaylor}), the kinetic equations in the non-dimensional form become,
\begin{align}
\left[\delta t' (\partial_{t'} + c'_{i\mu} \partial_{\mu'} ) + \frac{\delta t'^2}{2} (\partial_{t'} + c'_{i\mu} \partial_{\mu'} )^2\right] f_i &= { \omega_{\rm B} (f_i^{\rm eq} -f_i^{\rm B}) + \omega (f_i^{\rm B} -f_i) },
\label{eqn:ftaylorNonDimensional} 
\\
\left[\delta t' (\partial_{t'} + c'_{i\mu} \partial_{\mu'} ) + \frac{\delta t'^2}{2} (\partial_{t'} + c'_{i\mu} \partial_{\mu'} )^2\right] g_i &= \omega_1 (g_i^{\rm eq} -g_i) + (\omega - \omega_1) (g_i^* -g_i).
\label{eqn:gtaylorNonDimensional}
\end{align}
Let us define a smallness parameter $\epsilon$ as,
\begin{equation}
\epsilon=\delta t'=\frac{\delta t}{\bar t}.
\label{eqn:defineEpsillon} 
\end{equation}
Using the definition of $\epsilon$ and dropping the primes for ease of writing, we obtain,
\begin{align}
\left[\epsilon (\partial_t + c_{i\mu} \partial_\mu ) + \frac{\epsilon^2}{2} (\partial_t + c_{i\mu} \partial_\mu )^2\right] f_i &= { \omega_{\rm B} (f_i^{\rm eq} -f_i^{\rm B}) + \omega (f_i^{\rm B} -f_i) },
\label{eqn:feps} 
\\
\left[\epsilon (\partial_t + c_{i\mu} \partial_\mu ) + \frac{\epsilon^2}{2}(\partial_t + c_{i\mu} \partial_\mu )^2\right] g_i &= \omega_1 (g_i^{\rm eq} -g_i) + (\omega - \omega_1) (g_i^* -g_i).
\label{eqn:geps}
\end{align}
Writing a power series expansion in $\epsilon$ as,
\begin{align}
\partial_t &= \partial_t^{(1)} + \epsilon \partial_t^{(2)},
\label{eqn:epst} 
\\
f_{i} &= f_{i}^{(0)} + \epsilon f_{i}^{(1)} + \epsilon^2 f_{i}^{(2)},
\label{eqn:epsf}
\\
{f_{i}^{\rm B}} & {= f_{i}^{\rm B (0)} + \epsilon f_{i}^{\rm B (1)} + \epsilon^2 f_{i}^{\rm B (2)},}
\label{eqn:epsfB} 
\\
g_{i} &= g_{i}^{(0)} + \epsilon g_{i}^{(1)} + \epsilon^2 g_{i}^{(2)},
\label{eqn:epsg} 
\\
g_{i}^* &= g_{i}^{*(0)} + \epsilon g_{i}^{*(1)} + \epsilon^2 g_{i}^{*(2)},
\label{eqn:epsgstar} 
\end{align}
we substitute the equations (\ref{eqn:epst}--\ref{eqn:epsgstar}) into (\ref{eqn:feps}) and (\ref{eqn:geps}), and proceed with collecting terms of same order. This procedure is standard \citep{chapman_mathematical_1970}; for the specific case of the two-population LBM see, e.\ g., \citep{karlin_consistent_2013}. At order $\epsilon^0$, we get,
\begin{align}
f_i^{(0)} & = { f_{i}^{\rm B (0)} } = f_i^{\rm eq},
\label{eqn:f0feq} 
\\
g_i^{(0)} &= g_i^{*(0)} = g_i^{\rm eq}.
\label{eqn:g0geq} 
\end{align}
At order $\epsilon^1$, upon summation over the discrete velocities, we find,
\begin{align}
&\partial_t^{(1)}  \phi \rho + \partial_{\alpha} j_\alpha^{\rm eq} = 0,
\label{eqn:dt1rhoeps1}
\\
&\partial_t^{(1)} j_\alpha^{\rm eq} + \partial_\beta P_{\alpha \beta}^{\rm eq} = 0,
\label{eqn:dt1ueps1}
\\
&\partial_t^{(1)} ( \phi \rho E) + \partial_{\alpha} q_\alpha^{\rm eq} = 0.
\label{eqn:dt1Teps1}
\end{align}
Here, $\rho$ is the density of the fluid given by the zeroth moment of the $f$-populations in equation (\ref{eqn:fdensity}), $j_\alpha^{\rm eq}$ is the equilibrium momentum of the fluid as defined by equation (\ref{eqn:f1momMomentum}), $P_{\alpha \beta}^{\rm eq}$ is the equilibrium pressure tensor and $q_\alpha^{\rm eq}$ is the equilibrium heat flux as defined by equations (\ref{eq:Peq}) and (\ref{eqn:geq1mom}), respectively, and  $\rho E$ is the total energy of the fluid calculated as the zeroth moment of $g$-populations using equation (\ref{eq:totalE}).
Finally, at order $\epsilon^2$ we arrive at,
\begin{align}
&\partial_t^{(2)} \rho = 0,
\label{eqn:dt2rhoeps2}\\
%
&\partial_t^{(2)} j_\alpha^{\rm eq} + \partial_\beta \left( \frac{1}{2} - \frac{1}{\omega}  \right) (\partial_t^{(1)} P_{\alpha \beta}^{\rm eq} + \partial_\gamma Q_{\alpha \beta \gamma}^{\rm eq}) +{ \partial_\beta \left( 1- \frac{\omega_{\rm B}}{\omega} \right) P_{\alpha \beta}^{\rm B (1)} }= { j_\alpha^{\rm B (2)} (\omega - \omega_{\rm B}) },
\label{eqn:dt2ueps2}\\
%
&\partial_t^{(2)} ( \phi \rho E )+ \partial_\alpha \left[ \left( \frac{1}{2} - \frac{1}{\omega}\right) (\partial_t^{(1)} q_\alpha^{\rm eq} + \partial_\beta R_{\alpha \beta}^{\rm eq}) + \left( 1 -\frac{\omega_1}{\omega}\right) q_\alpha^{*(1)} \right] =0.
\label{eqn:dt2Teps2}
\end{align}
Here, $Q_{\alpha \beta \gamma}^{\rm eq}=\phi \rho \frac{u_\alpha}{\phi} \frac{u_\beta}{\phi} \frac{u_\gamma}{\phi} +  \phi P \left( \frac{u_\alpha}{\phi} \delta_{\beta \gamma} + \frac{u_\beta}{\phi} \delta_{\alpha \gamma} + \frac{u_\gamma}{\phi} \delta_{\alpha \beta} \right)$ is third-order moment of $f_i^{\rm eq}$ and $R_{\alpha \beta}^{\rm eq}$ is the second-order moment of $g_i^{\rm eq}$ given by (\ref{eqn:geq2mom}).
Combining terms at both orders, we recover the following macroscopic equations,
\begin{align}
&\partial_t  \phi \rho + \partial_{\alpha} j_\alpha^{\rm eq}=0, 
\label{eqn:dtRhoAppendix}\\
&\partial_t j_\alpha^{\rm eq} + \partial_\beta P_{\alpha \beta}^{\rm eq} 
 + \epsilon \partial_\beta \left( \frac{1}{2} - \frac{1}{\omega}  \right) (\partial_t^{(1)} P_{\alpha \beta}^{\rm eq} + \partial_\gamma Q_{\alpha \beta \gamma}^{\rm eq}) \nonumber \\ 
  & +{ \epsilon \partial_\beta \left( 1- \frac{\omega_{\rm B}}{\omega} \right) P_{\alpha \beta}^{\rm B (1)} }= { \epsilon j_\alpha^{\rm B (2)} (\omega - \omega_{\rm B}) },
\label{eqn:dtRhoUAppendix}\\
&\partial_t ( \phi \rho E) + \partial_{\alpha} q_\alpha^{\rm eq} + 
\epsilon  \partial_\alpha \left[ \left( \frac{1}{2} - \frac{1}{\omega}\right) (\partial_t^{(1)} q_\alpha^{\rm eq} + \partial_\beta R_{\alpha \beta}^{\rm eq}) + \left( 1 -\frac{\omega_1}{\omega}\right) q_\alpha^{*(1)} \right]  =0,
\label{eqn:dtRhoEAppendix}
\end{align}
where, 
\begin{align}
&\partial_t^{(1)} P_{\alpha \beta}^{\rm eq} + \partial_\gamma Q_{\alpha \beta \gamma}^{\rm eq} = P \left[ (\partial_\alpha u_\beta + \partial_\beta u_\alpha) + \left( \frac{2}{D}- \frac{R}{C_v} \right) \partial_\gamma u_\gamma \delta_{\alpha \beta} - \frac{2}{D} \partial_\gamma u_\gamma \delta_{\alpha \beta} \right], 
\label{eqn:dt1P} \\
&\partial_t^{(1)} q_\alpha^{\rm eq} + \partial_\beta R_{\alpha \beta}^{\rm eq} = \frac{u_\beta}{ \phi} P \left[ (\partial_\alpha u_\beta + \partial_\beta u_\alpha) + \left( \frac{2}{D}- \frac{R}{C_v} \right) \partial_\gamma u_\gamma \delta_{\alpha \beta} - \frac{2}{D} \partial_\gamma u_\gamma \delta_{\alpha \beta} \right] \nonumber \\
&{+ \phi P \sum_{a=1}^M H_a\partial_{\alpha}Y_a + \phi P C_p\partial_{\alpha}T,}
\label{eqn:dt1q} \\
&q_\alpha^{*(1)}= \left( \frac{1}{\omega_1} \right) (\partial_t^{(1)} q_\alpha^{\rm eq} + \partial_\beta R_{\alpha \beta}^{\rm eq}) + \frac{1}{\epsilon} \left( \frac{\omega}{\omega_1} \right) \left(- \frac{u_\beta}{ \phi} (P_{\alpha \beta}-P_{\alpha \beta}^{\rm eq}) + q_\alpha^{\rm diff} + q_\alpha^{\rm corr}+ {q_\alpha^{\rm B}} \right) ,
\label{eqn:qStar1} \\
 & P_{\alpha \beta}-P_{\alpha \beta}^{\rm eq} = \epsilon \left(-\frac{1}{\omega}\right) (\partial_t^{(1)} P_{\alpha \beta}^{\rm eq} + \partial_\gamma Q_{\alpha \beta \gamma}^{\rm eq}) + { \epsilon \left( 1 - \frac{\omega_{\rm B}}{\omega} \right) P_{\alpha \beta}^{\rm B (1)} } .
\label{eqn:Pneq}
\end{align}
We now substitute for the moments from the expressions (\ref{eqn:dt1P}) to (\ref{eqn:Pneq}) in equations (\ref{eqn:dtRhoAppendix}) to (\ref{eqn:dtRhoEAppendix}) and for the equilibrium moments to get the resulting macroscopic equations.
Equation (\ref{eqn:dtRhoAppendix}) recovers the continuity equation,
\begin{align}
&\partial_t  \phi \rho + \partial_\alpha (\rho u_\alpha) = 0.
\label{eqn:continuityMixAppendix} 
\end{align}
Equation (\ref{eqn:dtRhoUAppendix}) recovers the mixture momentum equation,
\begin{align}
&\partial_t (\rho u_\alpha) + {\frac{1}{\phi}} \partial_\beta (\rho u_\alpha u_\beta) + \partial_\beta \pi_{\alpha \beta}= {\mathcal{F}_\alpha}^{\rm k}, 
\label{eqn:momentumMixAppendix} 
\end{align}
with the constitutive relation for the stress tensor,
\begin{align}
\pi_{\alpha \beta} =  \phi P \delta_{\alpha \beta} - \mu 
\left(\partial_\alpha u_\beta + \partial_\beta u_\alpha - \frac{2}{D} (\partial_\mu u_\mu) \delta_{\alpha \beta}\right) - {\varsigma} (\partial_\mu u_\mu) \delta_{\alpha \beta}.
\label{eqn:piMix}
\end{align}
The dynamic viscosity $\mu$ is related to the relaxation coefficient $\omega$ by equation (\ref{eq:mu}) and the bulk viscosity $\varsigma$ is an input as described in equation (\ref{eqn:Pex}).
Finally, equation (\ref{eqn:dtRhoEAppendix}) recovers the mixture energy equation,
%
%
%
\begin{align}
\label{eqn:energyMixIntermediateAppendix}
\partial_t ( \phi \rho E) + \partial_\alpha(\rho E u_\alpha)+ \frac{1}{ \phi} \partial_\alpha ( \pi_{\alpha \beta}u_\beta) +\partial_{\alpha}q_{\alpha}=0,
\end{align}
where the heat flux $\bm{q}$ has the following form,
\begin{align}
\label{eqn:energyFluxIntermediateAppendix}
q_{\alpha}=-  \phi \lambda\partial_\alpha T - {{ \epsilon P \left( \frac{1
}{\omega_1}- \frac{1}{2} \right)  \phi \sum_{a=1}^M H_a \partial_\alpha Y_a } }+ {\left( \frac{\omega}{\omega_1}-1 \right) q_\alpha^{\rm corr}} + \left( \frac{\omega}{\omega_1}-1 \right) q_\alpha^{\rm diff},
\end{align}
with  the thermal conductivity $\lambda$ defined by equation (\ref{eq:lambda}). 
We now chose $q_\alpha^{\rm corr}$ to cancel the spurious second term containing the gradient of {$Y_a$},
\begin{align}
q_\alpha^{\rm corr} = \frac{1}{2} \left( \frac{\omega_1-2}{\omega_1-\omega} \right) \epsilon  P  \phi {\sum_{a=1}^M H_a \partial_\alpha Y_a .}
\label{eqn:qcorr}
\end{align}
This is equivalent to Eq.\ (\ref{eq:corrFourier}). Finally, the inter-diffusion energy flux is introduced by choosing the last term $\bm{q}^{\rm diff}$ in (\ref{eqn:energyFluxIntermediateAppendix}) as,
\begin{align}
q_\alpha^{\rm diff} = \left( \frac{\omega_1}{\omega-\omega_1} \right)  \rho \sum_{a=1}^{M} H_a Y_a V_{a\alpha},
\label{eqn:qdiff} 
\end{align}
which is equivalent to Eq.\ (\ref{eq:interdiffusion}).
Substituting (\ref{eqn:qcorr}) and (\ref{eqn:qdiff}) into (\ref{eqn:energyFluxIntermediateAppendix}), we get the heat flux ${\bm{q}}$ in the energy equation (\ref{eqn:energyMixIntermediateAppendix}) as a combination of the Fourier law and the inter-diffusion energy flux due to diffusion of the species \citep{kee_chemically_2017,williams_combustion_1985,bird_transport_2007},
%
\begin{equation}
{q_\alpha} = -  \phi \lambda \partial_\alpha T +  \rho \sum_{a=1}^{M} H_a Y_a V_{a\alpha}.   
\label{eqn:qMixCE} 
\end{equation}

\bibliographystyle{jfm}
\bibliography{ReadOnlyReferences} 

\end{document}